# Coherent view of crystal chemistry and ab initio analyses of Pb(II) and Bi(III) Lone Pair in square planar coordination.

Samir F. Matar[1]* and Jean Galy[2].

[1] CNRS-ICMCB, Université de Bordeaux, F-33600 Pessac, France.

[2] CNRS-LCTS, Université de Bordeaux, F-33600 Pessac, France.

E mail adresses:

*Corresponding author: S. F. Matar  matar@icmcb-bordeaux.cnrs.fr; abouliess@gmail.com

J. Galy  galy@lcts-bordeaux1.fr; jdgaly@orange.fr

**This paper is dedicated to Sten Andersson**

# A B S T R A C T

The stereochemistry of $6s^2$ (E) lone pair of divalent Pb and trivalent Bi (PbII and BiIII designated by M*) in structurally related PbO, PbFX (X= Cl, Br, I), BiOX (X= F, Cl, Br, I) and $Bi_2NbO_5F$ is rationalized. The lone pair LP presence determined by its sphere of influence E, equal to those of oxygen or fluorine anions, was settled by its center then giving M*-E directions and distances. Detailed description of structural features of both elements in the title compounds characterized by $[PbEO]_n$ and $[BiEO]_n$ layers allowed to show the evolution of M*-E distance versus the changes with the square pyramidal SP coordination polyhedra. All are different, in red PbO one finds $\{PbEO_4E_4\}$ square antiprism, a $\{[Bi.E]O_4X_4X_{apical}\}$ monocapped square antiprism in PbFX and BiOX and $\{BiEO_4F_4\}$ square antiprism in $Bi_2NbO_5F$. To analyze the crystal chemistry results, the electronic structures of these compounds were calculated within density functional theory DFT. Real space analyses of electron localization illustrate a full volume development of the lone pair on PbII within $\{PbEO_4E_4\}$ in PbOE, $\{PbEF_4X_4\}$ in PbFXE and Bi(III) within $\{BiEO_4X_4\}$ square antiprisms, contrary to Bi(III) within $\{[Bi.E]O_4F_4F_{apical}\}$ monocapped square antiprism. Larger hardness (larger bulk modules B0) and band gap characterize BiOF versus PbO due to the presence of F which brings antibonding Bi-F interactions oppositely to mainly bonding Bi-O. In PbFX and BiOX series there is a systematic decrease of B0 with the increasing volume following the nature and size of *X* which is decreasingly electronegative and increasingly large. The electronic densities of states mirror these effects through the relative energy position and relative electronegativities of F/X and O/X leading to decrease the band gap.

# I    Introduction.

The ions exhibiting an $ns^2$ ($n = 2 - 6$) outer electronic configuration are often stereo chemically active. Such M ions are designated herein by M*. As early as 1940 Sidgwick and Powell [1] underlined the equal importance of bonding pairs (BP) and lone pairs (LP) of electrons. Gillespie and Nyholm [2] and Gillespie and Hargitaï [3] assumed that lone pairs are larger than bond pairs with the corollary that repulsion between electron pairs decreases in the order LP − LP > LP − BP > BP − BP. This way to describe structure is achieved in the "Valence-Shell Electron-Pair Repulsion (VSEPR) model of molecular geometry". Gillespie [4] suggested a model for the packing of three BP on one hand, and two BP and one LP on the other hand. This is represented in Fig. 1 a, b showing the repulsive effect of LP through the reduction of the bond angle to below 120°. Note that VSEPR theory has long been criticized for not being quantitative, and therefore limited to the generation of "crude", even though structurally accurate, molecular geometries of covalent molecules.

<div align="center" style="color:red">Here Fig. 1</div>

In the Solid State, Andersson et al. [5] and Galy et al. [6,7] showed that the lone pair, designated by E, completes effectively the coordination polyhedron around the nucleus of M* cation, occupying a volume of the same order of magnitude as $O^{2-}$ or $F^-$, often building with oxygen atoms a hexagonal or cubic close packing. A simple test indicates the presence of E by calculating the so-called reduced volume $V_r$ which consists in dividing the cell volume V by the number of oxygen atoms (or fluorine atoms) and including the lone pairs E. For instance, in α-, β- PbOE and $(Pb^{2+})_2Pb^{4+}O_4E_2$ one finds $V_r(O,E) = 19.4$, 19.2 and 21.2Å$^3$ respectively; interestingly these magnitudes compare well with α-PbO$_2$ which has no LP Pb(IV) and where $V_r(O) = 20.3$Å$^3$.

Therefore a different approach of M* coordination was proposed. For example in CN = 4+1 coordination, i.e. triangular bipyramid (TBP), taking into account that E is close to M* and that it exhibits a volume similar to $O^{2-}$ (or $F^-$) it becomes possible to extract E coordinates by building a TBP M*O$_4$E with d(E–O) ~ d(O–O) (Fig. 2). In such a construction the expansion of M*–O axial bonds compared to M*–O equatorial ones, always found in crystal structures, is directly in agreement with the postulate that LP-BP repulsion is stronger than BP-BP one. The OM*O equatorial (<120°) and axial (<180°) angles are squeezed (reduced) and M*–O axial bonds are larger than equatorial ones, the latter being somewhat protected by M* cation. Such construction can be extended to the square pyramid, another CN = 4+1 coordination and to CN = 3+1, tetrahedron. Then a description of several structures was developed confirming the presence of E, generally evidenced in the network by a large vacancy capable to welcome a sphere having the size of $O^{2-}$ or $F^-$ anions. Its virtual center was geometrically determined and adjusted taking into account the closest surrounding atoms. Therefore a different approach of M* coordination was proposed. For example in CN = 3+1 or CN = 4+1 coordination, i.e. tetrahedral (T), triangular bipyramidal (TBP) or square pyramidal (SP) geometries it becomes possible to extract E coordinates by building a T $M*O_3E$, TBP or SP $M*O_4E$ with E-O ~ O-O (Fig 2a,b,c). In such construction the O-M*-O angle is <109.7° in T, axial bonds $M*-O_{ax}$ are bigger than equatorial ones $M*-O_{eq}$ with $O_{eq}M*O_{eq}$ <120° and $O_{ax}M*O_{ax}$ <180° angles in TBP, always found in crystal structures. The OM*O angles are squeezed and M*-O axial bonds are larger than equatorial ones in TBP. Such construction is extended with similar observations in SP.

<p style="text-align:center; color:red;">Here Fig. 2</p>

In view of the huge number of compounds containing M* ions, the aim of the present work is to follow the evolution of E and therefore M*-E distance within a three-dimensional network by focusing essentially on compounds comprising two *p* elements of the 6$^{th}$ period, namely the $6s^2$ Pb(II) and Bi(III) in square pyramidal SP (Fig. 2c) $M*O_4E$ or $M*F_4E$ coordination with atomic surroundings of M* and E being modified in different ways. To do so, red PbO (litharge), BiOF and $Bi_2NbO_5F$ crystal structures [9-14] were selected as models for the present study and depicted in Fig. 3. Therefore PbO and BiOF cases were extended to BiOX and PbFX with X = Cl, Br, I (15-17), isostructural with BiOF, in order to appreciate the impact of such bigger cations on E localization and therefore to $Bi^{3+}$ coordination scheme. In this context, the chemical formulas are labeled with E lone pair as: PbOE, BiOXE, PbFXE and $Bi_2NbO_5FE_2$ in which E sphere of influence is rationalized and illustrated on one hand and the electronic band structures and bonding properties are examined within the quantum density functional theory (DFT) [18, 19] on the other hand.

<p style="text-align:center; color:red;">Here Fig. 3</p>

Within DFT framework the electronic structures of PbO and BiOF have been investigated elsewhere [20-22] with results stressing the trends of electronic structures in the two varieties of PbO on one hand and within the BiOX series the changes due to the chemical nature of X halogen on the other hand. Concerning $Bi_2NbO_5F$ crystal structure (cf. Figs. 1,2) the symmetry of the space group *I4/mmm* formerly established by X rays [12] and refined by neutron powder diffraction [13-14], was lowered to *Pca*2$_1$ space group [14] in order to account for ferroelectric properties recently investigated [23] within the local density approximation (LDA) to DFT [24]. Nevertheless for the purpose of focusing on the study of E $6s^2$ lone pair effects comparatively between title compounds, the calculations were carried out here within the initial *I4/mmm* space group and assuming a primitive *P* mode to account for the presence of O/F order.

## II    Stereochemistry of $6s^2$ lone pairs of Pb(II) and Bi(III).

Cell parameters, internal atomic coordinates and reduced volumes are summarized in Table 1 together with calculated ones resulting from full geometry optimization within VASP method (cf. section III). Cell reduced volumes $V_r$(O,F,E) are simply calculated as defined above [5,6]. The large $V_r$(O,F,E) magnitude of PbOE is an indication of a large expansion of E, contrary to the two other values for Bi compounds. In spite of the close magnitudes of the Bi based compounds, the smaller BiOFE volume $V_r$, is an indication for LP spatially constrained development (cf. Figs 7 and 8). Such $V_r$ calculation applied for compounds containing large X halogen atoms like BiOX indicate the volume of the [BiOXE] unit.

The BiOXE series with X = Cl, Br, I, isostructural with BiOFE, are characterized by the presence of typical $[BiO]_n$ layer having the same conformation as $[PbO]_n$ being investigated. It was tempting to establish a parallel with the analogous PbFXE series [25-27], isostructural with BiOXE ones and then to BiOFE, which exhibits $[PbF]_n$ layers analogous to $[BiO]_n$ ones. PbFXE crystal data are included in Table 1.

here include TABLE 1

Red PbOE form is built up by $PbO_4E$ square pyramids sharing edges and corners, making parallel $[PbOE]_n$ layers packed along [001]. A perspective view of the structure is given in Fig. 3. E situated at the apex of $PbO_4E$ SP shows a Pb-E = 0.73Å distance. In the subfigures representing the crystal structures E sphere of influence is represented by a pale blue ellipse and its center by a blue dot; the choice of an ellipse allows indicating the direction in which E exerts its steric influence.

Therefore this steric influence on the lattice network is based on:

   i-      The E sphere radius estimated at $R_E$ = 1.27Å,

   ii-     an E-O(F) distance in the range 2.6 -2.7Å to oxygen (fluorine) atoms coordinated to M* and

iii-    a determination of reduced E coordinates taking care of immediate environment in order to avoid inadequate distances (i.e. E-O, F or E < 2.3Å).

A detailed representation of the respective environments of Pb and Bi in SP coordination is provided in Fig. 4. In Fig. 4a (Pb) and 4c (Bi) M*E are in a square antiprism (SAP) whereas BiE in Fig. 4b is found in a monocapped square antiprism (MSAP) in which $F_{ap}$ and $F_a$ designate apical and equatorial fluorine atoms.

<p style="text-align:center;color:red;">Here Figure 4</p>

BiOF crystal structure x-ray determination has been performed on synthetic single crystals [10]. Later on, neutron diffraction studies allowed better definition of crystal network [11]. Worthy to note that BiOF was also discovered as a mineral called "zavaritskite". As shown in Fig. 2b, BiOFE network shows remarkable features with its $[BiOE]_n$ layers, parallel to (001) plane, analogous to $[PbOE]_n$ ones (Fig. 2a) with oxygen atoms forming again a perfect square lattice. Bi atoms are alternately up and down above the oxygen layer with slightly shorter Bi-O bonds versus Pb-O due to trivalent Bi ($Bi^{3+}$) versus divalent ($Pb^{2+}$), thus exhibiting some covalent character. From a structural point of view $[BiOE]_n$ layer is absolutely similar to $[Bi_2O_2E_2]_n$ layer of the Aurivillius phase $[Bi_2O_2E_2]NbO_3F$ [12-14]. $Bi^{3+}$ then exhibits the classical coordination CN = 4+1 in the form of a square pyramid $BiO_4E$. Therefore, in BiOFE, these $[BiOE]_n$ layers are separated by puckered fluorine double layers with a puckering angle of 90.2° (Fig. 1). These layers dramatically modify $Bi^{3+}$ stereochemistry compared to $Pb^{2+}$. Five Bi-F interactions are formed, the shortest one along [001], $Bi-F_{ap}$ = 2.752Å, and four other $Bi-F_a$ these four fluorine atoms making another square net parallel to (001) with F-F = 3.747Å. Finally [Bi.E] appear encapsulated in a monocapped square antiprism (MSAP) $[Bi.E]O_4Fa_4F_{ap}$ (Fig. 1). Note that $Bi-F_{ap}$ is slightly smaller than $Bi-F_a$ distances by 0.041Å in spite of the presence of E. Taking into account the network features, particularly the $Bi-F_{ap}$ and F-F distances, E has been localized at Bi-E = 0.50Å which is a limit for acceptable E-O and E-F distances. Worthy to note the important diminution of Bi-E compared to Pb-E separations, the $[Bi.E](O_4F_5)$ unit encapsulation in the regular MSAP $\{O_4F_5\}$ being responsible of this fact (Fig. 2).

In BiOFE the established bonding explains why O-O distances are smaller than in PbOE (2.791Å and 2.656Å respectively) due to the $BiE-F_a$ interaction and also, even weak, with $Bi-F_{ap}$ along [001]. These interactions pull out of oxygen plane Bi atoms and also reduce O–Bi–O angles compared to O–Pb–O (see Table 2). Such a fact explains why Bi–Bi intra layer plane distance is increased compared to PbOE case. [Bi.E]-$F_a$ interactions are also responsible for *a* cell contraction compared to PbOE ones, 3.796Å against 3.947Å.

Finally the 3+ charge of Bi associated with relativistic effect of the large Z numbers explains the complicated stereochemistry of Bi with a close center E making a large deformed [Bi.E]$^{3+}$ cation. Then the distortion of the electronic cloud around the nucleus leads to this wide MSAP [Bi.E]$O_4F_5$.

For the sake of comparison some crystal data (distances, bonds and angles) characterizing these structures are given in Table 1. Bi-$F_{ap}$ interaction through E volume is weak, the distance 2.921Å being affected by the presence of E between these atoms.

Owing to the difference between Pb and Bi oxidation states some bond distances are bigger in PbOE than in BiOFE as shown in Table 1 together with some others. Worthy to note that in BiOFE ∠ O-Bi-O angle value is smaller than ∠ O-Pb-O by ~ 3.4°, directly linked to LP - BP repulsion in synergy with a shortening of Bi-O bonds and Bi-E distances

To emphasize the particular influence of these fluorine layers between the double layer of LP in [$Bi_2O_2E_2$]$_n$ it is interesting to show the result of the introduction of a slab [$NbO_3F$]$_n$, of $SnF_4$ type, built up by [$NbO_2(OF)_2$] octahedra sharing equatorial corners (Fig. 3).

The resulting $Bi_2NbO_5F$ oxyfluoride is the prototype (n = 1) of a large series of compounds [$Bi_2O_2E_2$]$_n^{2+}$ [$A_{n-1}B_nO_{3n+1}$]$_n^{2-}$ developed by Aurivillius [12]. The full disorder between O and F anions has been simplified, oxygen anions being assigned in the equatorial positions and remaining oxygen and fluorine (OF) were equally distributed onto apices. As shown in Fig. 3 this [$NbO_3F$]$_n$ layer leaves full space for Bi lone pair along fourfold axis alike in PbOE. Therefore it remains a weak bonding Bi-(OF) = 2.862 Å with the apices of [$NbO_2(OF)_2$] octahedra, but in roughly similar geometric conditions as Pb-$E_a$ or Bi-$F_a$ in previous structures but with no barrier for E development. Then Bi-E = 0.63Å distance becomes closer to Pb-E = 0.73Å.

These structural comparisons allowed to appreciate the LP role in the network architecture by its sphere of influence which appears rather constant and by its adjustment via M*-E distance versus the cations, here $Pb^{2+}$ and $Bi^{3+}$ in three different configurations with M*-E = 0.73 Å, 0.50 Å and 0.63 Å respectively and Pb.E in open SAP [$O_4Ea_4$], [Bi.E] enclosed in a MSAP [$O_4F_5$] and Bi.E in an open SAP [$O_4(OF)_4$].

In this context it became of paramount importance to analyze in detail the electronic distributions in these structures to enlighten these conclusions. Based on electronic band structure calculations, they are developed in the forthcoming paragraphs.

At this point we introduce the theoretical framework which will allow merging together the crystal chemical analysis and the theoretical one.

## III      A brief survey of the computational framework

Among the numerous codes built around and within DFT [18, 19], two computational methods were used in this work in a complementary manner. The Vienna ab initio simulation package (VASP) code [28-29] allows geometry optimization and cohesive energy calculations based on the projector augmented wave (PAW) method [30]. The effects of exchange and correlation are accounted for with the generalized gradient approximation (GGA) scheme following Perdew, Burke and Ernzerhof (PBE) [31]. Preliminary LDA [24] based calculations led to largely underestimated volumes versus experiment; they are not considered herein. One of the major outcomes of the calculations in tight relation with the topic of the present work is the rationalization of the electron localization. This is achieved thanks to the electron localization function (ELF) introduced by Becke and Edgecomb [32] as initially devised for Hartree–Fock calculations. Later on, its extension to DFT methods was done by Savin et al. [33] as based on the kinetic energy in which the Pauli Exclusion Principle is included:

ELF = $(1+ \chi_\sigma^2)^{-1}$ with $0 \leq$ ELF $\leq 1$, i.e. it is a normalized function.

In this expression the ratio $\chi_\sigma = D_\sigma/D_\sigma^0$, where $D_\sigma = \tau_\sigma - \nabla s - \frac{1}{4} (\nabla \rho_\sigma)^2/\rho_\sigma$ and $D_\sigma^\circ = 3/5$ $(6\pi^2)^{2/3} \rho_\sigma^{5/3}$ correspond respectively to a measure of Pauli repulsion ($D_\sigma$) of the actual system and to the free electron gas repulsion ($D_\sigma^0$) and $\tau_\sigma$ is the kinetic energy density. Then a normalization of the ELF function between 0 (zero localization) and 1 (strong localization) with the value of ½ corresponding to a free electron gas behavior enables analyzing the contour plots following a color code: blue zones for zero localization, red zones for full localization and green zone for ELF= ½, corresponding to a free electron gas (cf. Fig. 8). Beside the 2D ELF representation we consider the corresponding 3D isosurfaces enclosing the electrons of each atomic constituent. It will be shown the usefulness of such representations for the discussion of the lone pair development and stereoactivity.

In a second step, all-electrons calculations, equally within the GGA were carried out for a full description of the electronic structure, particularly the site projected density of states (DOS), PDOS and the properties of chemical bonding. The augmented spherical wave (ASW) method was devised by Williams et al. [34] and further developed by Eyert continuously [35] leading to full potential FP-ASW with implementation of chemical bonding according to different schemes. Herein we discuss the pair interactions based on the overlap population analysis with the crystal orbital overlap population (COOP) [36]. In the plots, positive, negative, and zero COOP indicate bonding, anti-bonding, and non-bonding interactions, respectively.

For further details on the two DFT methods used here, the reader is kindly referred to a recent review [37].

**IV- The lone pair and its ELF representation in PbOE.**

A qualitative description of PbE in PbOE is provided in Fig. 5a, left hand side (LHS), showing the vertical plane containing O-Pb-O and E, together with corresponding ELF on the right hand side (RHS) projected along a vertical plane (101) (cf. Fig. 4a).

We firstly discuss the LHS sketch. The Pb–O distance d(Pb-O)= 2.304Å is reported. [PbE] are alternatively distributed above and below a perfect square net of oxygen atoms. This entity is developed inside a large square antiprism (SAP), $\{[PbE]O_4E_4\}$, formed by four oxygen atoms and four LP's of the following $[PbOE]_n$ layer (cf. Fig. 2). The layer thickness reaches 3.951Å while E-E interlayer distance amounts to 1.16Å. This shows that LP – LP interactions, which finally associate these $[PbO_4E]_n$ layers, constituting stable red PbO crystal or powder, could be compared to Van der Waals ones. Worthy to note the appropriateness of the geometric choice describing Pb.E stereochemistry: the E size is deducted from former structure analyses of various oxides and is close to the radius of $O^{2-}$, i.e. ~1.27Å. We also estimate O-E equal to 2.75Å. Then Pb-E is obtained from geometric construction after having determined the position of E along the tetragonal $c$ axis, i.e. along [001] direction. Consequently Pb-E distance is 0.73Å. This important value will be the criterion of network constraints exerted on Pb(II) LP. It is clear that E is the center of the LP sphere of influence.

Here Figure 5

After this description, a better data rationalizing should be brought by the ELF projection obtained from high accuracy self consistent calculations (cf Table 1). The corresponding 3D ELF isosurface shell shown in Fig. 5 RHS, brings further important features:

*i*- the PbOE shell exhibits an egg form like the postulated shape by geometric description (interpenetrated spheres of Pb and E). It is grossly enclosed in a sphere around 2.55Å in diameter (dotted blue circle - blue point center (slightly below E);

*ii*- The shell exhibits a ~2.85Å height and a maximum diameter of ~2.58Å against 2.94Å and 2.54Å (left); both values measuring its distortion;

*iii*- E is then settled in the middle of the large diameter of the shell which indicates the maximum expansion of the sphere of influence (different from blue circle center);

*iv*- d(Pb-E) = 0.65Å. Then E position is directly estimated onto shell picture and its reduced coordinates are derived;

*v*- the red circle (same size of the blue one i.e. 2.55Å) was centered on E and slightly deformed to fit the ELF shell. To do so it was elongated towards the oxygenated base of the square [PbO$_4$] pyramid testifying of LP-BP repulsion of E towards Pb–O bonding with electronic cloud deformation as a result. In PbOE, LP has full space to develop (E-E$_a$ = 3.15Å).

The inter-atomic distances in the crystal structure and in the ELF projection are compared in Table 1. This schematic view will find various illustrations and representations with the discussion of ELFs in the course of this paper.

## V- Geometry optimization and energy volume equation of state.

From Table 1, the unconstrained full geometry optimizations for all structures provide good agreement with experiment for the cell parameters, volumes and *z* coordinates. The larger discrepancy found for Bi$_2$NbO$_5$F is due to the symmetry breaking upon going from body center *I* mode of *I4/mmm* space group to the primitive *P* mode through operating the (½,½,½) translations leading to a doubling of the number of atoms needed to account for the occupation of the 4*e* (0,0,$z_{O/F}$) positions by 50% of O and F (cf. Figs. 2c and 4c). Nevertheless the agreement is good enough to obtain the ELF (cf. next section) and to derive trends in the physical properties of the title compounds pertaining to the mechanical ones, i.e. the response of the crystal lattice to external pressure. This is particularly relevant in view of the larger volume of BiOF versus PbO with a significantly larger *c/a* ratio: *c/a*(BiOF) = 1.66 versus *c/a*(PbO) = 1.18. For the sake of establishing mechanical trends related with the crystal chemistry, the whole series of PbFX then BiOX were explored.

Also in spite of the similarity of the $[BiOE]_n$ layer with the $[Bi_2O_2E_2]_n$ layer in $Bi_2NbO_5F$, it becomes important to identify the mechanical role played by the additional Nb-O-F motifs, i.e. with $[Bi_2O_2E_2][NbO_3F]$.

The mechanical properties, particularly those pertaining to the compressibility $B_0$ (bulk modulus), can be analyzed from the equilibrium zero pressure parameters derived from the energy-volume, E(V), equation of state (EOS).

$B_0$ magnitudes within a family of compounds ex. PbFX and BiOX, are particularly relevant for appreciating the role of the changing chemical element as well as the role of an extra atom in a structure, here regarding the changes with F from PbO to BiOF dominated by $[PbOE]_n$ and $[BiOE]_n$ layers.

The calculations around minima found from geometry optimization provide the E,V pair values and the resulting E = $f$(V) curves (Figs. 6). The fit of the curves with a Birch EOS [39]:

$$E(V) = E_o(V_o) + [9/8]V_oB_o[([(V_o)/V])^{2/3}-1]^2 + [9/16]B_o(B^{'}-4)V_o[([(V_o)/V])^{2/3}-1]^3,$$

with $E_o$, $V_o$, $B_o$ and $B^{'}$, designating the equilibrium energy, volume, bulk modulus and its pressure derivative respectively. The low $\chi^2$ magnitudes demonstrate the goodness of fit, i.e. the lower the $\chi^2$, the better the fit.

The fit parameters given in the inserts of the respective curves reproduce the trends of the geometry optimization for the volume with a better agreement of the volume with the experiment as with respect to geometry optimization in Table 1. The corresponding $B_0$ are also provided in the Table.

In Fig. 6a illustrating PbO and BiOF, the striking feature is in the changing magnitudes of the bulk modules: $B_0(PbO) = 80.5$ GPa and $B_0(BiOF) = 121$ GPa. They are significantly different and range below the magnitudes observed for oxides (~200 GPa) while approaching those usually found in (soft) alloys [40]. From the respective $B_0$ magnitudes PbO is a readily compressible oxide versus harder BiOF. But such trends are somehow opposite to the expected ones regarding volume considerations: the larger the volume, the more compressible the compound. Consequently the larger incompressibility of BiOF can be assigned to the presence of fluorine atoms in the interplanes of $[BiOE]_n$, on top of Bi.E, and Bi.E-F interaction should be repulsive as shown in next section through the COOP analysis. On the contrary the empty space between [PbOE]$_n$ layers leads to a largely compressible oxide (cf. Figs. 1).

Turning to the PbFX series, Fig. 6b shows the energy – volume curves for the three compounds with the fit values. The bulk modulus magnitude decreases with increasing volume due to the increasing size of X from Cl to Br and I. Here B$_0$ values are systematically lower than in PbO in spite of the presence of *X* on top of Pb.E. This can be assessed based on nature of anion bonded to Pb in the square planar pyramid, i.e. F versus O: the longer Pb–F bonds (~2.6Å) versus ~2.3Å for Pb-O have an effect which overrides the sterical effect of apical *X* (cf. Table 1).

The change of X in the BiOX series is discussed from the Birch EOS results in Fig. 6c. There is a systematic decrease of B$_0$ with the increasing volume along the series following the nature and size of *X* which is decreasingly electronegative and increasingly large.

Lastly we point out to the incompressibility of Bi$_2$NbO$_5$F with the largest magnitude B$_0$ = 157.0 GPa obtained among the studied compounds (Table 1). Clearly the additional [NbO$_3$F] motifs, adjoined to [Bi$_2$O$_2$E$_2$] ones have the role of decreasing the compressibility of the quaternary compound with respect to the above discussed compounds. Note that also at the atomic constituents' level, the bulk module of Nb ~170 GPa is much higher than that of Bi ~ 31 GPa.

B' pressure derivatives of B$_0$ magnitudes of all compounds are close to those usually found in the solid state [40].

<span style="color:red">Here Figs. 6)</span>

### VI. Electron localization function ELF

A- 2 D slices

Figs. 7 show the ELF slices –2D views–. The color legends shown by the ruler at the bottom of Figs. 7, i.e. blue, green and red ELF indicate areas corresponding to zero, free electron (ELF = ½) like and full localization (ELF = 1).

<span style="background-color:yellow">Here Figs. 7 revoir texte car il n'y a plus 7b-e et f...</span>

The ELF slice of PbOE is reproduced in the first panel. Its discussion follows the detailed description in section V above. In PbF*X*, the introduction of *X* on top of Pb.E can be seen to modify the ELF around Pb by visually making closer the spacing between the centers of LP and Pb. This is further detailed and quantified in next section pertaining to the 3D ELF shells.

Also the breadth of the LP is seen to decrease along the series. The bonding between the different constituents, i.e. Pb–F(O) on one hand and Pb –X on the other hand is illustrated by the green areas. In the first kind of bonding in Pb–O is visually stronger than Pb–F as materialized by broader and less localized green areas; notice the appearance of blue zones of zero localization underneath the F–Pb–F ELF. The Pb–X ELFs show decreasing green interaction zones which disappear in PbFI. Simultaneously with the increase of c/a tetragonality ratio, the ELF around Pb tends to be elongated towards X and less towards F, *viz.* in the panel of PbFI. This involves changes of bonding magnitudes discussed in next paragraphs pertaining to the chemical bonding.

The BiOX series exhibits similar O-Bi-O interaction green zones and large changes along Bi.E –$X_{ap.}$ whereby the distortion of Bi.E is largest for BiOFE and decreases significantly along the series with the onset of blue zero localization zones as soon as X changes to Cl. Note that in the cases of BiOBr and BiOI the same effect of elongation of Pb ELF towards *X*, i.e. as in PbF*X*, is concomitant with less localization between Bi and O. This also has consequences on the bonding (cf. Figs 10).

The ELF features presented by the quaternary compound complete our understanding of the trends above by showing the full development of the Bi LP within the apical space of the *BiEO$_4$F$_4$* square antiprism, free from atomic occupation, depicted in Fig. 4c.

Lastly we note that the presented electron localization picture is not that of an ionic like cation – anion one but corresponds more to covalent PbO and iono-covalent BiOF, the larger ionic property being brought by fluorine –this is no more valid for the other halogens-. In this context we suggest that the peculiar behavior of BiOF can be related with the property of ionic conductivity and the fluorine ion mobility formerly studied by us experimentally in the oxyfluoride as well as in solid solutions of BiOF with fluorite-type β-PbF$_2$ [41,42].

**Here Figs 8**

B- 3D ELF shells

Using crystallographic data, stereochemical interpretation and 3D ELF shell calculation applied to PbOE it was shown the remarkable agreement between the two approaches (see Fig. 5). The center E of the so-called LP volume of influence was localized in the middle of the maximum diameter of the shell (2.58Å), perpendicular to [001] in a [Pb.E] shell which shows an "egg" shape. The shell height along [001] and the diameter amount 2.85Å and 2.58Å respectively. Two major informations were extracted: the distances O-E = 2.75Å and Pb-E = 0.65Å. They are in good agreement with the postulated O-E distance value applied as based on O-O distance and after E geometric localization giving Pb-E (0.73Å) to the crystal structure. Worthy to note that all distances implying E are given at ± ~0.01Å.

Last point, it appears that the joint influence of E lone pair and Pb-O bonding explain the slimming of the shell toward the oxygen plane while the following $E_a$ layer offer plenty of space for E LP development on the opposite side ($E_a$-$E_a$ = 3.947Å forming a square plane above [Pb.E] shell).

On this "strategic" base the same analyses were performed successively onto BiOFE and $Bi_2NbO_5FE_2$. The ELF 3D shells of both compounds are shown in Fig. 8a.

Analyzing the former one it is shown that the whole shell can be grossly inserted into a sphere of 2.40Å in diameter (blue dotted circle). Therefore the [Bi.E] shell included in $\{O_4Fa_4Fap\}$ MSAP shows a main flattening due to $F_{ap}$ repulsive effect on its volume which maintains a large size of 2.37Å in spite of weak Bi- $F_a$ interactions. The height of [Bi.E] shell is reduced to 2.31Å giving this "pear" shape instead of "egg" one. E being settled, Bi-E = 0.47Å and O-E = 2.60Å distances were derived. This result is quite different compared to PbOE case. Obviously such [Bi.E] ELF shape is linked to respective neighboring of [Pb.E] and [Bi.E]: an open $\{O_4Ea_4\}$ SAP for the former and a MSAP for the second. Bi-O bonds with more covalent character have also diminished O-O distance by 5.1% which impacts the O-E distance by a similar value. The $F_a$ square plane (Bi- $F_a$ = 2.809Å) cuts the shell volume and $F_{ap}$ (Bi- $F_{ap}$ = 2.765Å) on the top achieve to constrain the shell.

BiOFE shell reveals the remarkable versatility of the LP ELF volume capable to adapt its shape to the constraints of the network, therefore keeping its basic volume.

This important result needed to be immediately supported by an example in which the [Bi.E] ELF shell was liberated from the capping of the $\{O_4,F_4\}$SAP. Such situation is achieved in $Bi_2NbO_5FE_2$ crystal structure in which the substitution of $[NbO_2(OF)_2]_n$ layer for the double $\{F\}_n$ layer of BiOF between the $[BiO]_n$ layers is realized. Then E can extend, a situation enlightened in Fig.8a where both 3D ELF shapes are exhibited for sake of comparison.

**HERE Fig 8a**

If again [Bi.E] shell shows a volume corresponding to a sphere of grossly ~2.4Å in diameter, now its shape comes back to an "egg" form alike in PbOE (Fig. 5). Worthy to note its height which amounts to ~2.76Å, a value slightly smaller by 0.09Å, therefore considerably higher than in BiOFE reaching only 2.31Å. We also note that if [Bi.E] shell top emerges from the (OF) square face of its {O$_4$(OF)$_4$}SAP it falls in an empty electron blue zone. E being readily set up, it appears that O-E =2.60Å and the important Bi-E distance compared to BiOFE one increases from 2.47Å up to 2.76Å. Starting from Bi the shape of the shell electronic cloud is also elongated towards the oxygen plane up to 1.13Å against 1.05Å in former case.

The choice of O-E distance to set up E in a crystal structure determines its position. In these two Bi compounds Bi-E value was the same i.e. 2.60Å, smaller than in PbOE case where it amounts to 2.75Å. It then became important to check if a reasonable constant value could be defined, even slightly different, for Pb as well as for Bi, keeping always in mind the O-O (or F-F) distances of the square base M*O$_4$E SP typical coordination scheme.

To achieve this task it was decided to follow this evolution in both series of compounds PbFXE and BiOXE with X=Cl, Br and I. These series exhibit the same space group and are isostructural with BiOFE as shown by their crystallographic data reported in Table 1.

The pictures of the PbFXE ELFs and BiOX ELFs reported in Fig. 8b and Fig 8c were then carefully analyzed and compared to the former PbOE ELF in Fig. 5 on one hand and BiOFE ELF in Fig 8a on the other hand.

HERE FIG 8b

In PbFXE series the [PbF]$_n$ layers show some differences notably a square net plane with F-F distances > 2.9Å somewhat larger than in the oxygen atoms network of PbOE (2.791Å). Within [PbF$_4$] square pyramids Pb-F average bond 2.55Å is also markedly larger than Pb-O bond of 2.30Å testifying certain ionic character.

Therefore [Pb.E] sits now inside a monocapped square antiprism [F$_4$X$_{a_4}$X$_{ap}$], alike Bi in BiOFE (MSAP [O$_4$F$_{a_4}$F$_{ap}$]), and the ELF shell is exposed to repulsion and constraints of the double layer of halogens Cl, Br or I intercalated in between [PbF]$_n$ layers. But here [Pb.E] shells are quasi perfectly enclosed in a sphere of radius ~2.5Å marked by a dotted blue circle alike in PbOE.

In PbFClE the ELF shell does not exhibit the same flattening as in BiOF, the chlorine anion being at Pb-Cl$_{ap}$ = 3.178Å against Bi- F$_{ap}$ = 2.765Å. Anyhow the shell appears rather close to a sphere even slightly made thinner towards the F base because of Pb-F and PbCl$_a$ bonds interactions. Worthy to note also that Cl$_a$ plane makes a limit to PbE shell expansion along [001]. In PbFBrE, the *c* parameter has increased, with enlarged distances Pb-Br$_{ap}$ and Pb-Br$_a$ and more slightly Pb-F bonds; Br$_a$ plane seems to be again a limit for PbE shell expansion therefore the height of the shell is now 2.65Å instead of 2.52Å previously. In both compounds Pb-E are very close 0.23Å and 0.21Å and the calculated E-F distance amounts 2.70Å. Iodine brings another big steps in $c$ parameter expansion Pb-I$_{ap}$ = 4.393Å some 0.944Å more than for Br and the I$_a$ plane is now above the [Pb.E] shell. So, note its large expansion which amounts 2.80Å while its largest diameter is similar to Br case (2.49Å) then forming an egg shape. Pb-E distance becomes smaller, 0.17Å, but F-E = 2.70Å like in former cases.

To conclude, it comes that if the fluorine base of MSAP has increased together with Pb-X$_a$ and Pb-X$_{ap}$, E sites being settled on diameter center, the distance E-F remains roughly the same (i.e. 2.70Å) but Pb-E separation has dramatically decreased compared to PbOE case (0.65Å); this is also observed with Bi-E in BiOFE case (0.47Å). In these large cells the lone pair E becomes increasingly closer to the cation center Pb.

In BiOXE series with X= F, Cl, Br and I the [BiO]$_n$ layers are characterized by Bi-O bonds which evolves from 2.296Å up to 2.398 with X = F to I and remains clearly smaller than Pb-F bonds in PbFXE series. The [Bi.E] ELF shells are somewhat smaller with diameters of LP sphere of influence around 2.40Å which roughly encloses them. Therefore their heights vary from 2.31Å for BiOF up to 2.54Å for BiOIE with an intermediate value 2.45Å for both Cl and Br anions.

HERE FIG 8c

If in BiOFE the [Bi.E] shell is clearly squeezed by F$_{ap}$ (Bi- F$_{ap}$ = 2.765Å) giving the pear shape to the shell and limiting the height of {O$_4$Fa$_4$} SAP at 2.254Å, the BiOXE's with remaining halogens show markedly different features, which is associated with Bi-X$_{ap}$ lengthening. With BiOClE the cell volume already increases dramatically (+ 130%), a blue space (zero localization) appears between [Bi.E] shell (which comes back to an egg shape) and the height of [Bi.E]O$_4$Cl$_4$ SAP becomes 2.592Å. Bi-E diminishes to 0.38Å, closer respective centers. The square Cl$_a$ plane is now onto the limit of the shell. Of course these features are again enlarged with Br and I, with the result to see the [Bi.E] shell closely tending to spherical shape, M*-E decreasing to 0.36Å for the former and to 0.27Å for the latter. The {[Bi.E]O$_4$I$_{a_4}$} SAP exhibits a height of 3.084Å and Bi-I$_{ap}$ = 4.823Å, a so large value that it cannot be consider as capping this SAP.

It appears again that when the atomic surrounding becomes wide or broad enough, the tendency for M*-E is to diminish and to build a large $[Bi.E]^{3+}$ spherical cation like a big $Ba^{2+}$ or $K^+$ then losing a large part of its one sided stereochemical activity.

## VII- Electronic structure and chemical bonding

The above partial observations should be complemented with an account of the electronic structures and chemical bonding properties.

In so far that the calculated structure parameters are close to experiment for the title compounds (Table 1) we used the latter for the analysis of the electronic band structure with all electrons scalar relativistic ASW method [20,21]. For the present atomic species, in the minimal basis set of the ASW method the valence states and the matrix elements were constructed using partial waves up to $l_{max}+1 = 3$ for Bi, Nb and Pb and $l_{max}+1 = 2$ for O and F; $l$ being the secondary quantum number. Also the low energy lying and filled sub shells of $F(2s^2)$, $O(2s^2)$ as well as Bi and Pb $5d^{10}$ were not considered as part of the valence basis sets and replaced by higher level corresponding empty states, thus contributing better to the completeness of the valence basis set.

HERE Figs 9

A       Site projected density of states DOS (PDOS)

Figures 9 present the site projected DOS (PDOS) of all compounds studied in this work. PbO is characterized by a broad and continuous valence band (VB) separated from a conduction band (CB) by a gap of ~1.7 eV. The PDOS features and gap magnitude are close to those obtained by Payne et al. [22]. The top of the VB is dominated by Pb and O $p$ states and the lower part comprises $s$ like states. All partial DOS show similar shape signaling their quantum mixing, leading to the chemical bonding detailed in next section. The broad features of the PbO PDOS are in agreement with the $[PbO]_n$ layered structure revealing its covalent character, in opposition to the PbFX series discussed below. PbFX series (X = Cl, Br, I) at Figs. 9b – 9d) show significant differences with PbO, within the VB, the band gap as well as the CB:

i-    The halogen at the apical position above Pb, has systematically lower electronegativity χ than F: χ(F) = 3.98> χ(Cl) = 3.16 > χ(Br) = 2.96 > χ(I) = 2.66. Then F $2p$ states are localized at lower energy than X. Specifically the F $2p$ states within the VB are centered at ~-3 eV in PbFCl and PbFBr, but they are found centered at lower energy at −6 eV in the case of PbFI, i.e. they are further stabilized. This is concomitant with the significantly different ELF shape of PbFI versus PbFCl and PbFBr at Fig. 7 where blue zero localization zones separate I from F-Pb-F layer. Also Pb–F distance increases in the series (Table 1).

ii-    The band gap decreases along the series from ~3.8 eV (X = Cl) to ~2.5 eV (X = Br) then down to ~1.6 eV in PbFI. This is again due to the chemical nature (less electronegative) and size of X which increases significantly up to I and to the separation of the I-ELF from the F-Pb-F ELF (cf. Fig. 7).

iii-    The lower part of the VB in all three panels shows similar shapes of the PDOS for all three constituents, in agreement with the quantum mixing involving $p$ as well as lower energy lying $s$ states.

The BiOX PDOS at Figs. 9e – 9h) show global features resembling those of PbFX, i.e. for the positions of $s$ states at the lower part of the VB, which is then dominated by $p$ states at the top on one hand and to the decreasing band gap magnitude from X = F to I on the other hand. Furthermore, the series offers another illustration of the chemical role of X in changing the electronic structure: the electronegativity values $\chi$ given above. This sequence explains the relative positions of O and X $p$ states: In BiOF the close magnitude of O and F electronegativities leads to close PDOS centering respectively at ~ -2 eV and ~-2 .4 eV. With X= Cl which is less electronegative than O, Cl-$p$ states are found at the top of VB. This is also observed with slightly larger separation in BiOBr and a clear separation between the $p$ states of O and I as identified for BiOI. Interestingly this feature is concomitant with the clear separation of the I-ELF from the basal $[BiO]_n$ in Fig. 7 as well as with the closing of the band gap as in PbFI.

In $Bi_2NbO_5F$, the large covalence due to 5/1 O - F ratio leads to the small band gap observed in Fig. 9i. The top of the VB is dominated by oxygen $p$ states and F $p$ states are found centered around -5 eV.

B      Chemical bonding from the overlap integrals analyses with COOP criterion.

The effect of the quantum mixing between the valence states leads to the chemical bonding. The analysis of the overlap population is based on the overlap integral $S_{ij}$ where i and j designate two chemical species with their secondary quantum numbers corresponding to $s$, $p$ or $d$ states. Such analysis is done with the so called 'crystal orbital overlap population' COOP [37], implemented in the within the ASW method [36]. Here we focus of the directional $p$ states in as far as all the chemical constituents are known as $p$-elements belonging to 3A, 4A, 6A and 7A columns of the periodic table of the elements.

<div style="text-align:center; color:red;">Here Figs. 10</div>

Figs. 10 show the different COOP plots in which positive, negative and zero $y$-axis magnitudes designate bonding, antibonding and non bonding interactions.

Panel a) shows the Pb–O interaction in PbO which is of mainly bonding character especially in the energy window -5 – -2 eV pertaining to Pb $p(x,y)$ and less intensity for the higher energy lying $p_z$ at ~-1 eV. For PbF$X$ family, panels 10b – 10d show the changes in the structure with the presence of apical X on top of Pb. Pb–F bond shows similar features with Pb–O especially for the lower energy part pertaining to the $p(x,y)$ orbitals. Towards the top of VB bonding Pb–X COOP develop oppositely to antibonding Pb–F. Most remarkable features are observed for PbFI (Fig. 10d) where dominant bonding interaction is for Pb–I. This is concomitant with the ELF feature at Fig. 7 where the Pb-ELF develops towards I with less expansion in the plane. The electronegativity difference between F and X increases from Cl to I and the series becomes somehow less ionic. Also the Pb–X distance increases leading to less antibonding interactions.

With respect to PbO, the major change in the BiOX series, as pointed out above, is in the presence of apical X. The effect mirroring the ELF slices (Figs. 7) and shells (Fig. 8) where the close Bi-F separation leads to the large distortion of the Bi and well as F ELFs, is observed for BiOF at Fig. 10e. The bonding Bi–O COOP are accompanied with bonding Bi-F in the energy window -5, -2 eV. Above this energy range largely antibonding Bi-F COOP are identified and the top of VB is antibonding. In BiOCl the shift of Bi–O to lower energy is observed while Bi-Cl bonding COOPs are shifted towards upper energy. The VB top is mainly of Bi–Cl antibonding character. The increase of $c$ tetragonal parameter from Cl to Br and I, concomitantly with the increase of $a$ parameter leads to

   i- A decreasing antibonding character (as in PbFX), and in BiOI the top of VB is dominated by bonding Bi–I COOP involving the Bi $p_z$ orbitals while the lower energy part of the VB shows the O–Bi–O bonding within [BiO]$_n$ involving $p_{x,y}$.

This latter aspect is less visible in PbFI where Pb–F bonds are much less visible in the lower energy part of VB.

ii-    A decreasing bonding character for Bi-O along the increase of a parameter.

For $Bi_2NbO_5F$ (Fig. 10i) the VB is dominated by Bi-O and Nb-O bonds and Bi-O COOP have similarities with those of BiOF with the difference of much less significant antibonding Bi-F COOP. This is concomitant with the development of Bi LP without distortion as observed in the ELF at both Figs. 7 and 8. Lastly in all panels the CB is dominated by antibonding counterparts

**Concluding remarks**

In this work we have shown that the virtual character of crystal chemistry approach of E lone pair can be rationalized and illustrated further with the support of DFT based quantum calculations.

Then the size and centering of the lone pair E volume has been refined:

i-    if the reduced volume $V_r$ (calculated taking into account the number of O, F and LP without cation volume contribution) is devoted to E, it comes from ELFs that in fact it corresponds to the shell [M*.E] volume (here above [Pb.E] and [Bi.E]) in the case of CN 4+1 one sided coordination to O or F of these elements;

ii-    this [M*.E] volume, which roughly equals a sphere, exhibits in fact an "egg" shape with the largest size opposite to one sided M*-O (or F) bonding. E has been localized on the centre of this pseudo diameter allowing to appreciate two important distances, i.e. O(F)-E and M*-E. The halogen at the apical position above Pb, has systematically lower influence on [Pb.E] ELF shape versus increasing its Z number.

iii-    the study of BiOFE has revealed another important phenomenon that is the flexibility of E shape versus its atomic surrounding. The fluorine of the intercalated double F layer between $[BiO]_n$ layers flattens the ELF egg shape until it gives it a pear form. M*-E increases if space is freed, a situation encountered in and well illustrated by $Bi_2NbO_5FE_2$. When the constraint of the network increases, surrounding anions coming close to E, the associate deformation in the case of one sided coordination follows M*-E diminution. Therefore, when the surrounding network is created by largest anions there is enough space for M*-E expansion but it also diminishes, a fact well illustrated with X = Cl, Br and I. In the case of PbFXE, [Pb.E] ELF shape becomes almost spherical with a particularly low Pb-E = 0.17Å value.

iv-    The E center of LP volume of influence, localized using an average value of O-O (or F-F) base of the SP by using a distance O-E = O-O, must be settled according to values revealed by the ELFs, i.e. ~2.70Å for [$PbO_4E$] or [$PbF_4E$] and 2.55-2.60Å for [$BiO_4E$]. <mark>Then as shown in Table 1 the M*- E values are in pretty good agreement for both crystal and ELF data.</mark> Worthy to note that along the studied compounds O-E or F-E do not evolve with the determined distances O-O or F-F distances of square bases which show increased values with X atoms network content.

Clearly the present work opens a broad horizon <mark>of for</mark> future investigations under development, pertaining to:

i-   The assessment of $6s^2$ lone pair in other environments as tetrahedral (T) and triangular bipyramid (TBP) –cf. fig. 2.

ii-   The extension of the studies of the $ns^2$ to smaller $n$ values, i.e. to light elements, comparatively in a given column (ex. IV-B; VI-B …);

iii- and, *in fine,* the assessment of the behavior of $E_c$ defining the centroïd of LP, responsible of the remarkable [M*E] ELF shells evolutions.

***Acknowledgements***. The authors express their thanks to Pr. Jean Etourneau for organizing a fruitful meeting completed by stimulating scientific discussions. Computations were conducted on MCIA-University of Bordeaux computer center.

Supports from CNRS and *Conseil Régional d'Aquitaine* are gratefully acknowledged.

**Table 1 –**

Experimental and (calculated) crystal data of red PbOE, PbFXE, BiOFE, BiOXE and $Bi_2NbO_5FE_2$ (E designating the lone pair and X Cl, Br or I). All these compounds crystallize in tetragonal system, with the space groups *P4/nmm* (origin 1) for the PbOE to BiOXE with O and F of the square nets are at (2*a*), Bi, Pb and F at (2*c*) and $Bi_2NbO_5F$ with *I*4/*mmm* space group with Nb at 2(a), Bi and O/F at 4(e), O1 at 4(c) and O2 at 4(d). $V_{EOS}$ and $B_0(GPa)$ refer to the equilibrium volume and the bulk modulus obtained from equation of state (EOS) fit using Birch EOS.

| Compounds | a (Å) | c (Å) | $z$ | V (Å³) | $V_{EOS}$ (Å³) | $B_0(GPa)$ | $V_\tau$ (O,F,E) (Å³) |
|---|---|---|---|---|---|---|---|
| Red PbOE [9] | 3.947 | 4.988 | $z_{Pb}$ 0.238(0.241) | 77.7 | | | 19.4 |
| | (4.06) | (4.994) | | (82.3) | 81.54 | 80.5 | |
| BiOFE [10,11] | 3.756 | 6.234 | $z_{Bi}$ 0.2077(0.209) | 87.4 | | | 14.9 |
| | (3.750) | (6.341) | $z_F$ 0.6524(0.645) | (89.2) | 89.69 | 121 | |
| $Bi_2NbO_5FE_2$[12] | 3.8348 | 16.64 | $z_{Bi}$ 0.3261(0.310) | 244.7 | | | 15.3 |
| | (3.900) | (16.75) | $z_{OF}$ 0.1174(0.12) | (254.8) | 247.64 | 157.0 | |
| | | | | | | | $V_\tau$ (M*(O,F)XE)(Å³) |
| BiOClE [13] | 3.887 | 7.354 | $z_{Bi}$ 0.1714(0.172) | 111.1 | | | 18.6 |
| | (3.890) | (7.363) | $z_{Cl}$ 0.6459(0.648) | (111.4) | 119.37 | 88.7 | |
| BiOBrE [14] | 3.923 | 8.105 | $z_{Bi}$ 0.154(0.150) | 124.8 | | | 21.1 |
| | (3.940) | (8.170) | $z_{Br}$ 0.657(0.660) | (126.8) | 128.44 | 82.0 | |
| BiOIE [15] | 3.995 | 9.151 | $z_{Bi}$ 0.1338(0.136) | 146.1 | | | 24.3 |
| | (4.10) | (9.151) | $z_I$ 0.6671(0.663) | (153.8) | 149.45 | 69.3 | |
| PbFClE | 4.1062 | 7.2264 | $z_{Pb}$ 0.2055(0.209) | 121.8 | | | 20.3 |
| | (4.11 ) | (7.29) | $z_{Cl}$ 0.6485(0.645) | (123.1) | 124.47 | 57.6 | |
| PbFBrE | 4.180 | 7.590 | $z_{Pb}$ 0.195(0.194) | 132.6 | | | 22.1 |
| | (4.21) | (7.63) | $z_{Br}$ 0.6500( 0.646) | (135.2) | 137.41 | 53.3 | |
| PbFIE | 4.2374 | 8.800 | $z_{Pb}$ 0.1639(0.169) | 158.0 | | | 26.3 |
| | (4.208) | (9.02 ) | $z_I$ 0.6630(0.656) | (160.1) | 158.95 | 43.9 | |

Interatomic distances (Å) and angles (°) from X-ray or neutron crystal structures (M* = Bi or Pb and X = F, Cl, Br or I). E is geometrically located based on E-O distances approximated from O–O distances or from ELFs 3D projections.

| M*OXE | M*-O | O-O | M*-X$_a$, (OF),E$_a$ | M*X$_{ap}$ | X$_a$-X$_{ap}$ | OM*O | E-O | M*-E | E-X$_a$ (OF),E$_a$ |
|---|---|---|---|---|---|---|---|---|---|
| PbOE | 2.304 | 2.791 | 3.37 | | | 117.8 | 2.75 | 0.73 | 2.98 |
| PbOE ELF | 2.360 | 2.871 | 3.46 | | | 118.7 | 2.75 | 0.65 | 3.15 |
| BiOFE | 2.281 | 2.656 | 2.795 | 2.772 | 3.266 | 110.8 | 2.60 | 0.50 | 2.68 |
| BiOFE ELF | 2.296 | 2.652 | 2.809 | 2.765 | 3.227 | 109.5 | 2.60 | 0.47 | 2.69 |
| Bi$_2$NbO$_5$FE | 2.298 | 2.712 | 2.870 | | | 113.1 | 2.70 | 0.63 | 2.73 |
| Bi$_2$NbO$_5$FE ELF | 2.194 | 2.717 | 2.997 | | | 125.5 | 2.60 | 0.71 | 2.80 |
| BiOClE | 2.317 | 2.749 | 3.059 | 3.490 | 3.487 | 114.1 | 2.60 | 0.47 | 2.89 |
| BiOClE ELF | 2.321 | 2.751 | 3.053 | 3.505 | 3.509 | 113.9 | 2.55 | 0.38 | 2.91 |
| BiOBrE | 2.325 | 2.774 | 3.170 | 4.075 | 3.763 | 115.1 | 2.60 | 0.46 | 2.98 |
| BiOBrE ELF | 2.380 | 2.786 | 3.185 | 4.110 | 3.787 | 114.9 | 2.55 | 0.36 | 3.03 |
| BiOIE | 2.343 | 2.825 | 3.362 | 4.880 | 4.163 | 117.0 | 2.60 | 0.44 | 3.14 |
| BiOIE ELF | 2.398 | 2.899 | 3.433 | 4.823 | 4.160 | 117.5 | 2.55 | 0.27 | 3.30 |

| M*FXE | M*-F | F-F | M*-Xa | M*Xap | Xa-Xap | FM*F | E-F | M*-E | E-Xa |
|---|---|---|---|---|---|---|---|---|---|
| PbFCl | 2.534 | 2.904 | 3.089 | 3.201 | 3.611 | 108.2 | 2.70 | 0.27 | 3.01 |
| PbFCl ELF | 2.558 | 2.906 | 3.095 | 3.178 | 3.594 | 106.9 | 2.70 | 0.23 | 3.02 |
| PbFBr | 2.561 | 2.956 | 3.181 | 3.453 | 3.731 | 109.4 | 2.70 | 0.23 | 3.10 |
| PbFBr ELF | 2.573 | 2.977 | 3.218 | 3.449 | 3.718 | 109.8 | 2.70 | 0.21 | 3.13 |
| PbFI | 2.563 | 2.996 | 3.361 | 4.392 | 4.148 | 111.5 | 2.70 | 0.23 | 3.26 |
| PbFI ELF | 2.598 | 2.976 | 3.368 | 4.393 | 4.096 | 108.2 | 2.70 | 0.17 | 3.29 |

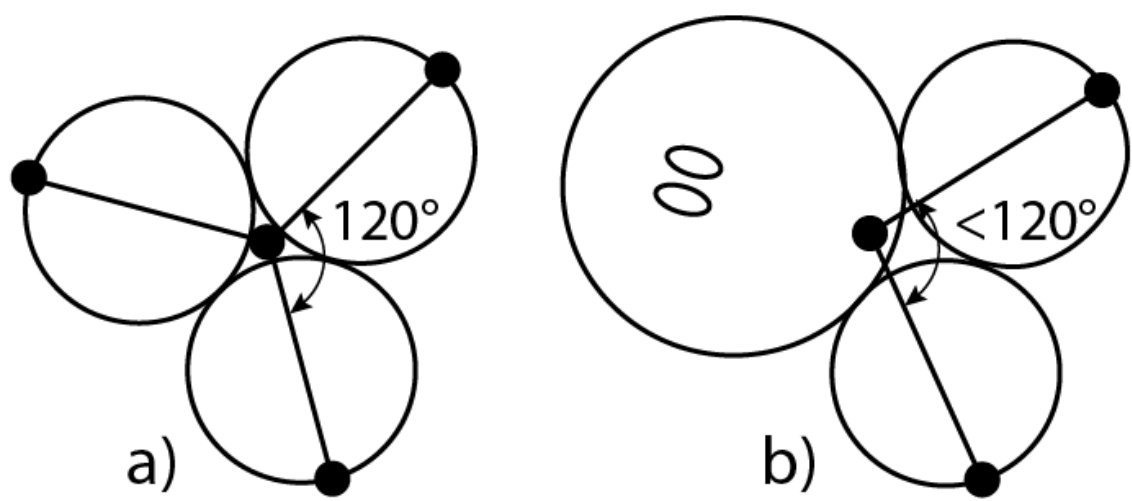

**Figure 1** Gillespie model for bond pairs BP and lone pairs LP: Small black circles represent the electron core of atoms, large circles stand for BP and the circle with two small ellipses represents the LP: a) Trigonal packing; b) LP repulsion acts by enlarging atomic distances and by corollary squeezing the angle to below 120°.

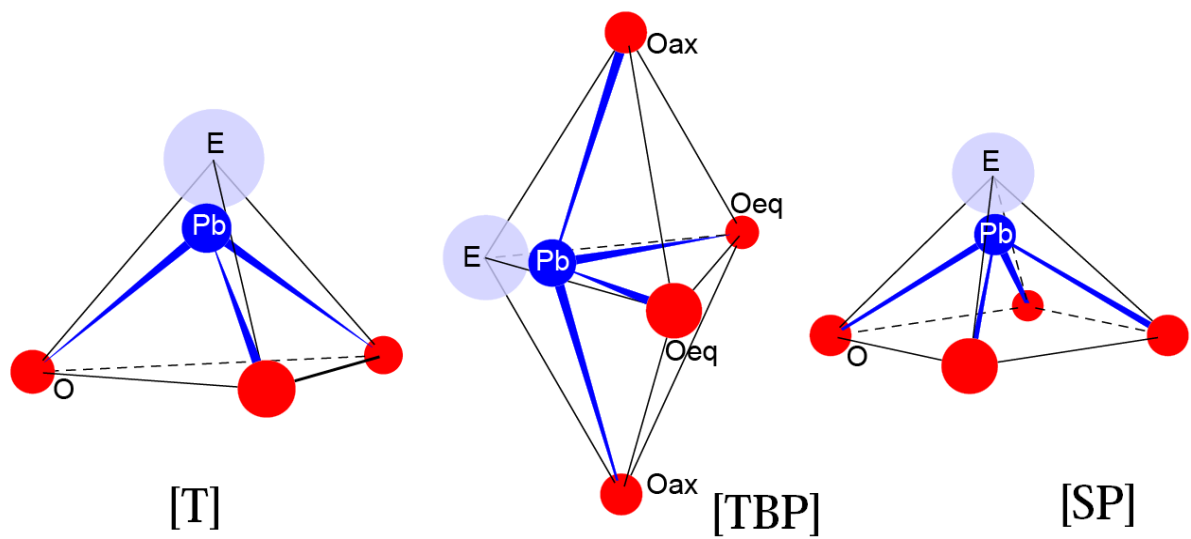

**Figure 2.** Geometry of three coordination polyhedral types around M* element, here Pb for tetrahedral [T], triangular bipyramidal [TBP] and square pyramidal [SP].
Note: By swinging E in TBP towards one $O_{ax}$, T tetrahedron is obtained and if $O_{ax}M*O_{ax}$ angle diminishes TBP tends towards SP.

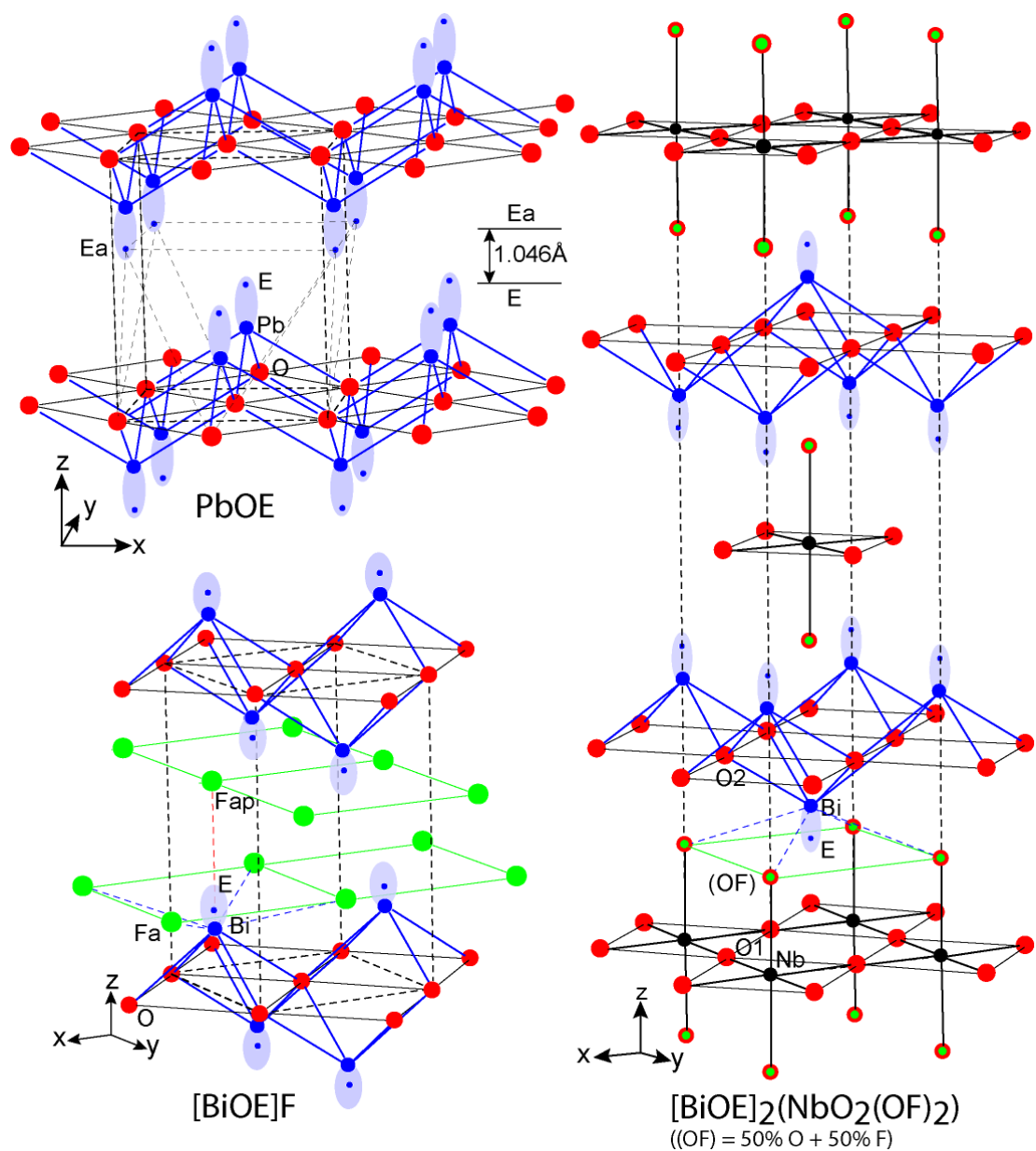

**Figure 3** (color online) PbOE - [PbO$_4$E] square pyramid and [O$_4$E$_{a4}$] square antiprism around PbE (E represents the lone pair). [PbOE]$_n$ layers are packed along [001] and "associated" via a double layer of lone pairs. BiOFE - [BiOE]$_n$ layers, isostructural with [PbOE]$_n$ ones, are now separated by a double layer of fluorine atoms just intercalated and BiE sits inside a monocapped [O$_4$F$_{a4}$F$_{ap}$] square antiprism. F$_{ap}$ is exactly in Bi-E direction (dotted red line) (atoms: Bi blue, O red, F green, E pale blue). Bi$_2$NbO$_5$FE$_2$ - [Bi$_2$O$_2$E$_2$]$_n$ layer is closely related to those of PbOE and BiOFE but the oxygen square net is slightly smaller than the former and bigger than the second (see Table 1).

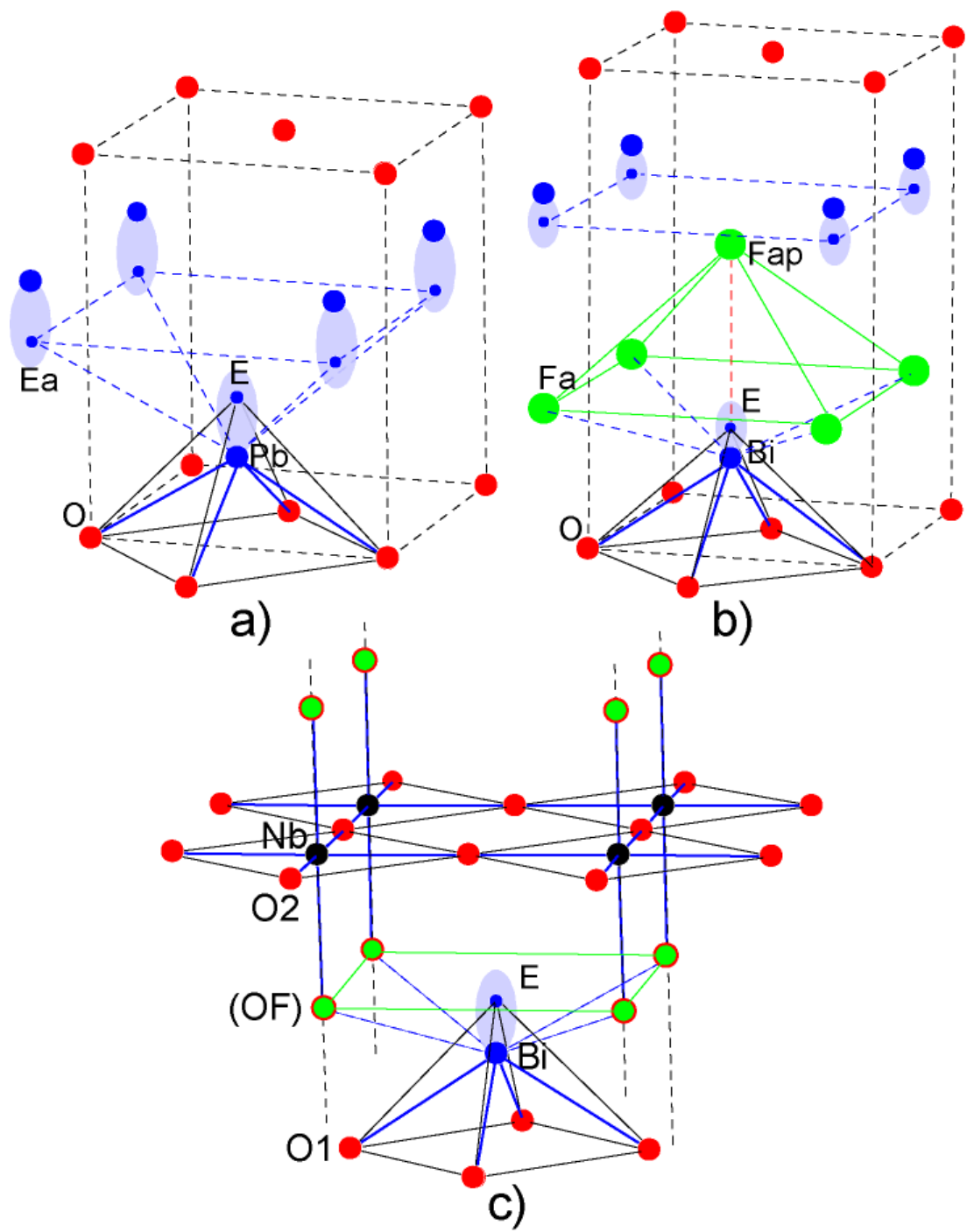

**Figure 4** (color) Comparison of PbE and BiE environments in PbOE, BiOFE and Bi$_2$NbO$_5$FE$_2$: a) PbEO$_4$E$_{a_4}$ square antiprism ; b) [Bi.E]O$_4$F$_{a_4}$F$_{ap}$ monocapped square antiprism; c) BiEO$_4$F$_4$ square antiprism.

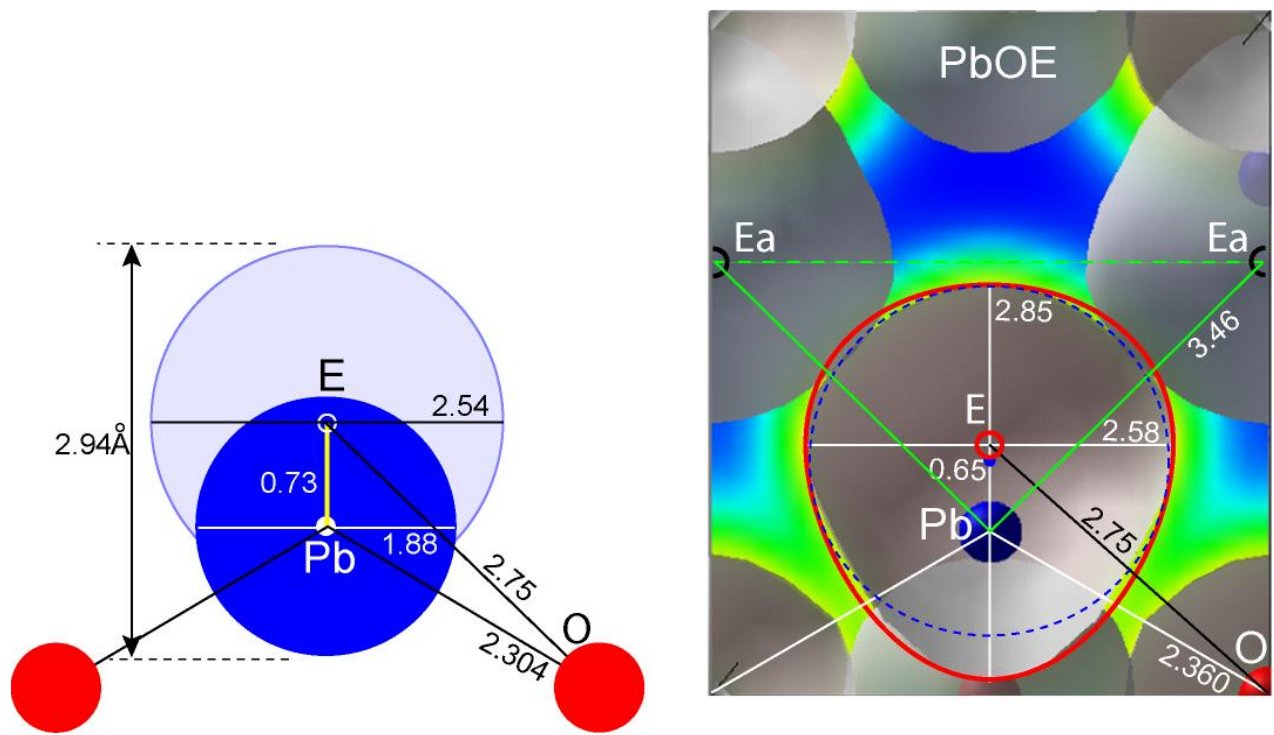

**Figure 5** (color) Geometric deduction of E lone pair center from crystal data (left) and inside [PbE] shell issued from ELF calculation (right).

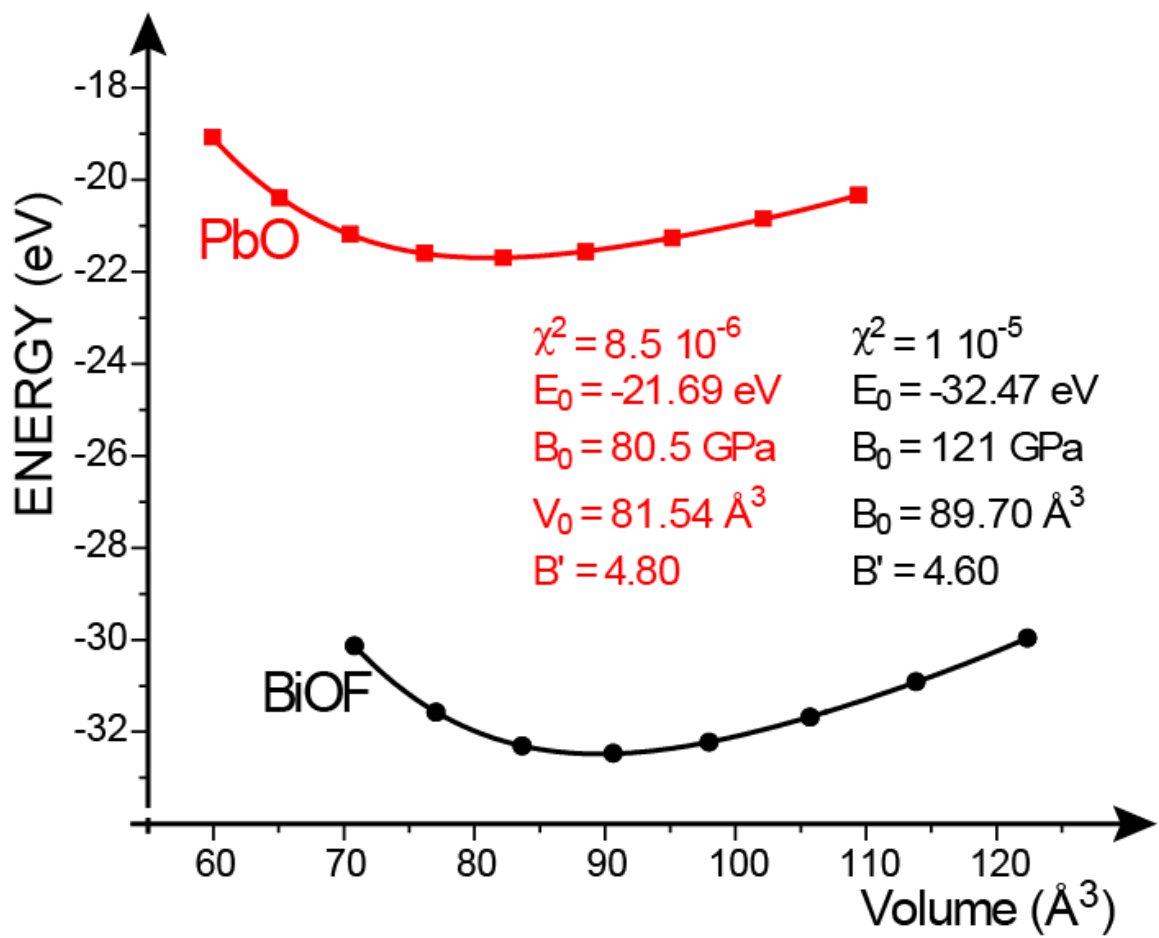

a)

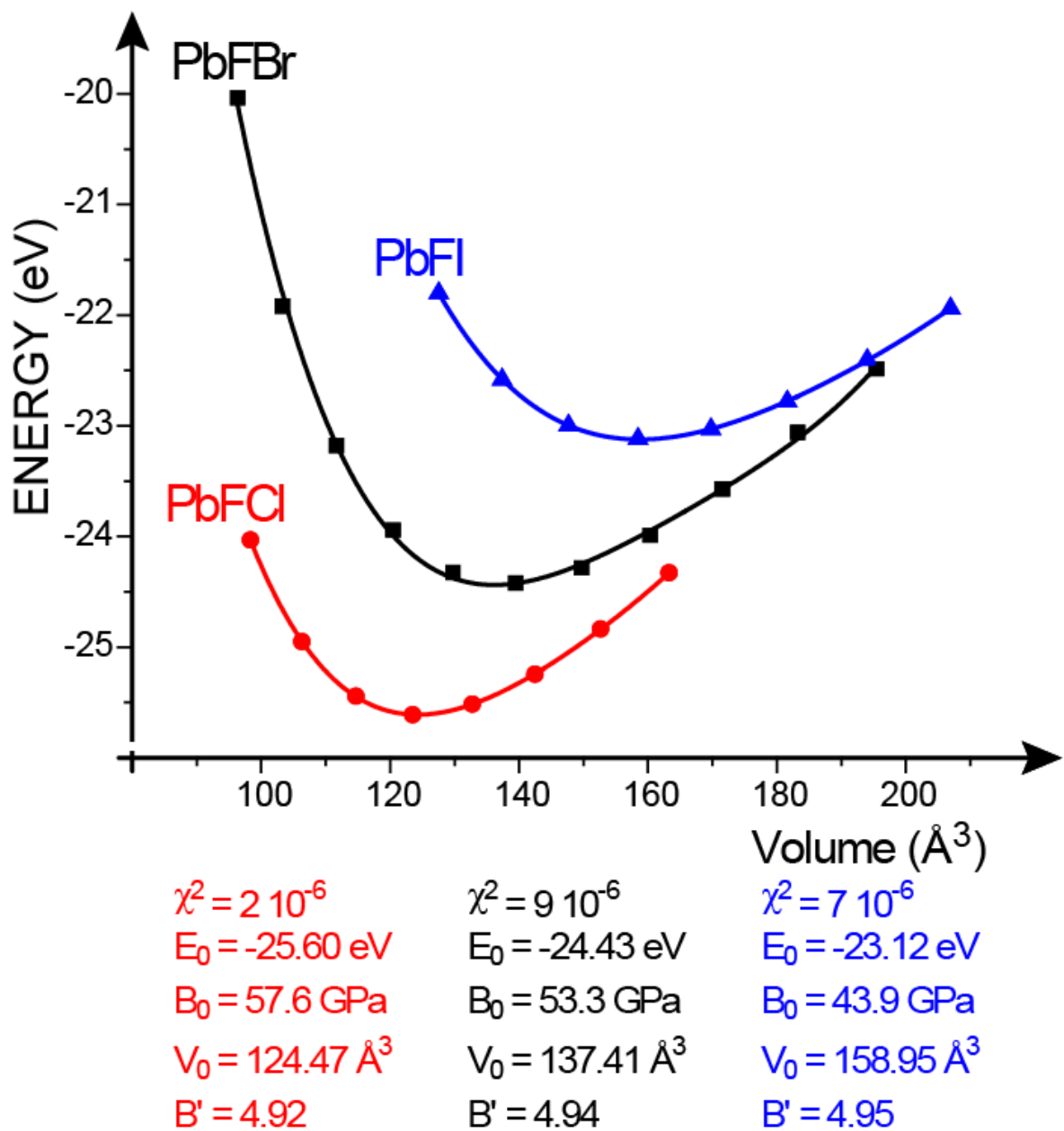

$\chi^2 = 2\ 10^{-6}$     $\chi^2 = 9\ 10^{-6}$     $\chi^2 = 7\ 10^{-6}$
$E_0 = -25.60$ eV     $E_0 = -24.43$ eV     $E_0 = -23.12$ eV
$B_0 = 57.6$ GPa     $B_0 = 53.3$ GPa     $B_0 = 43.9$ GPa
$V_0 = 124.47$ Å$^3$     $V_0 = 137.41$ Å$^3$     $V_0 = 158.95$ Å$^3$
$B' = 4.92$     $B' = 4.94$     $B' = 4.95$

b)

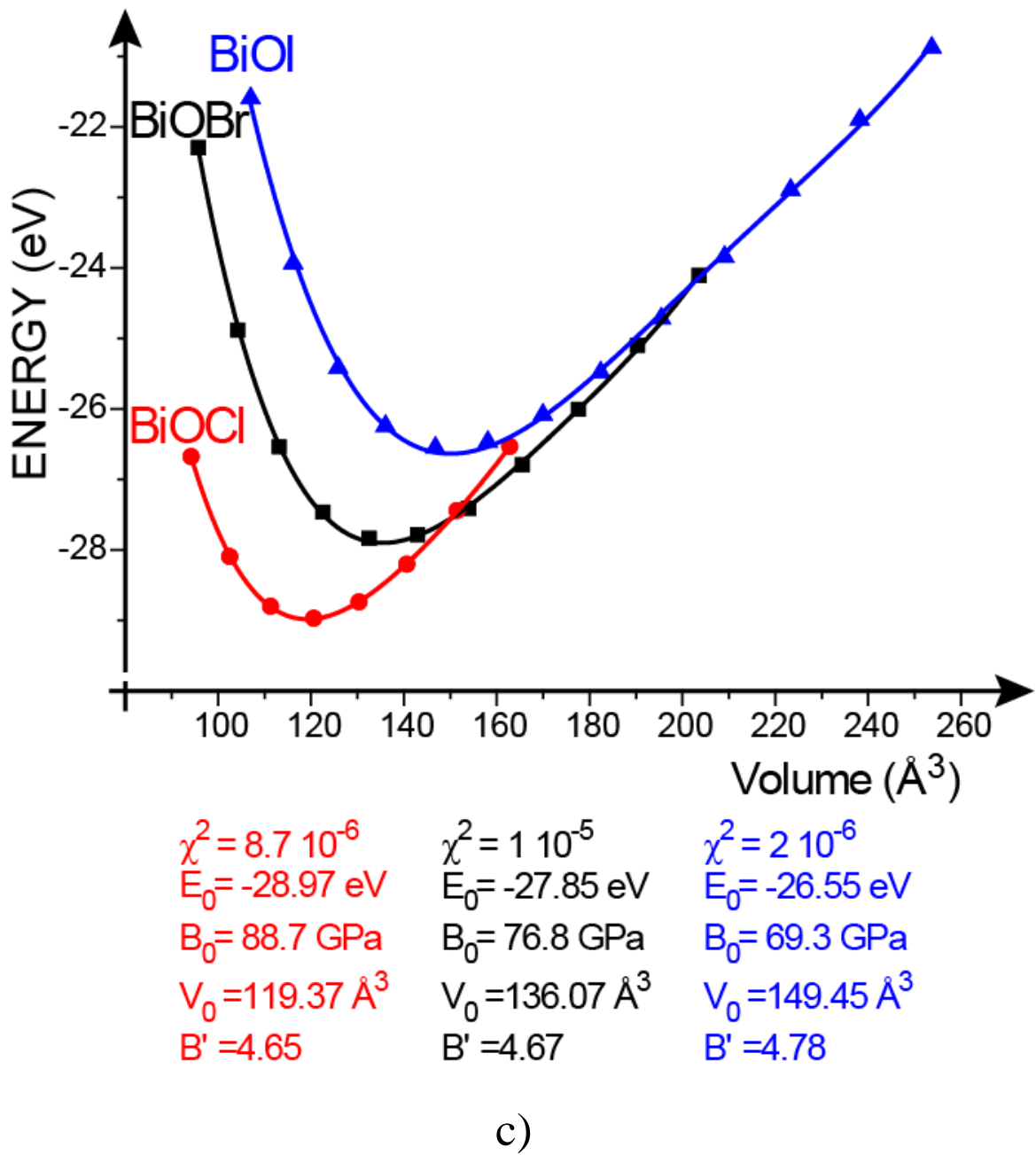

**Fig. 6.** (color) Energy versus volume curves of a) PbO and BiOF, b) PbFX series and c) BiOX series. Fit parameters of energy ($E_0$), bulk modulus ($B_0$) and volume ($V_0$) are obtained from Birch $3^{rd}$ order equation of state (EOS). $\chi^2$ magnitudes signal the goodness of fit.

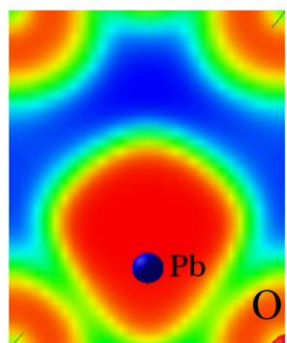

PbOE

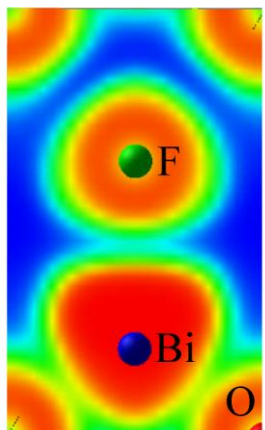

BiOFE

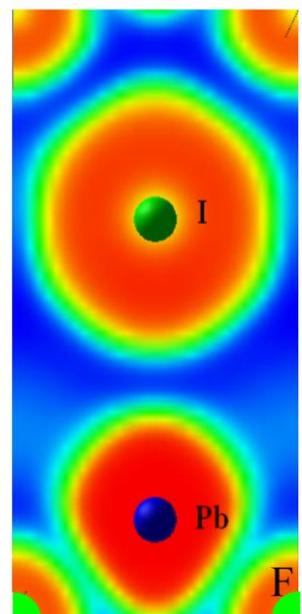

PbFIE

Bi$_2$NbO$_5$FE$_2$

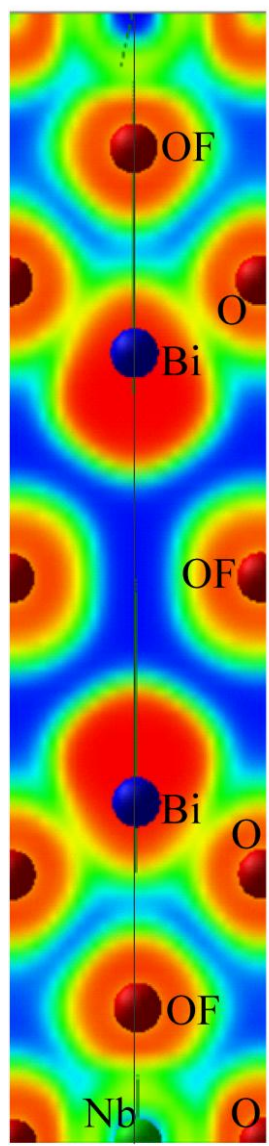

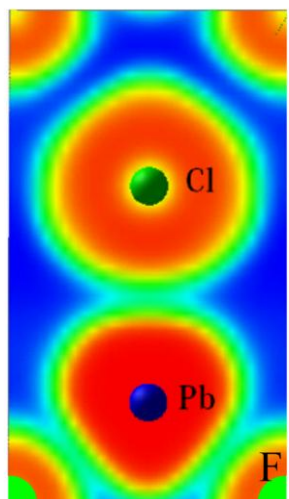

PbFClE

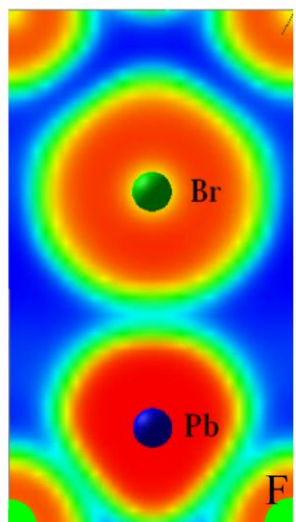

PbFBrE

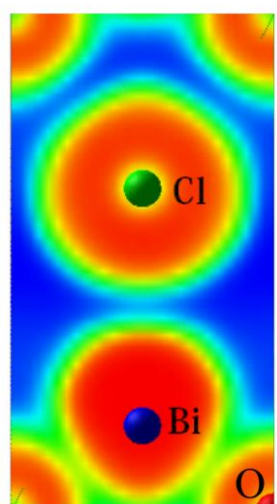

BiOClE

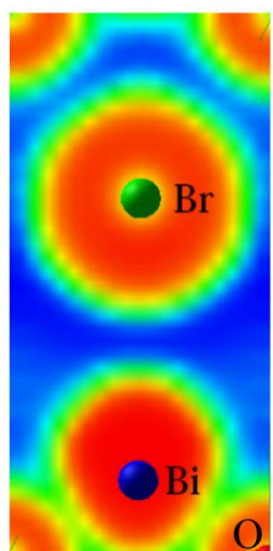

BiOBrE

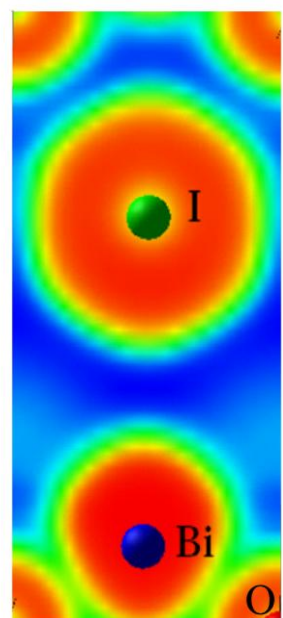

BiOIE

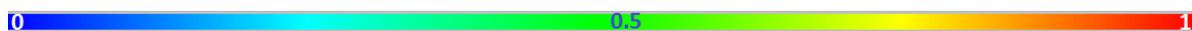

**Figure 7** (color) Electron localization function ELF slices along (100) tetragonal plane of PbO, BiOX (X = F, Cl, Br, I) and $Bi_2NbO_5F$. The rular shows the color code with normalized ELF values bewteen 0 (no localization) and 1 (full localization).

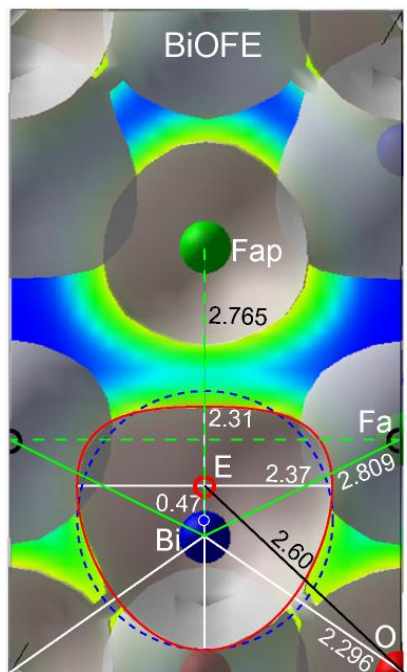

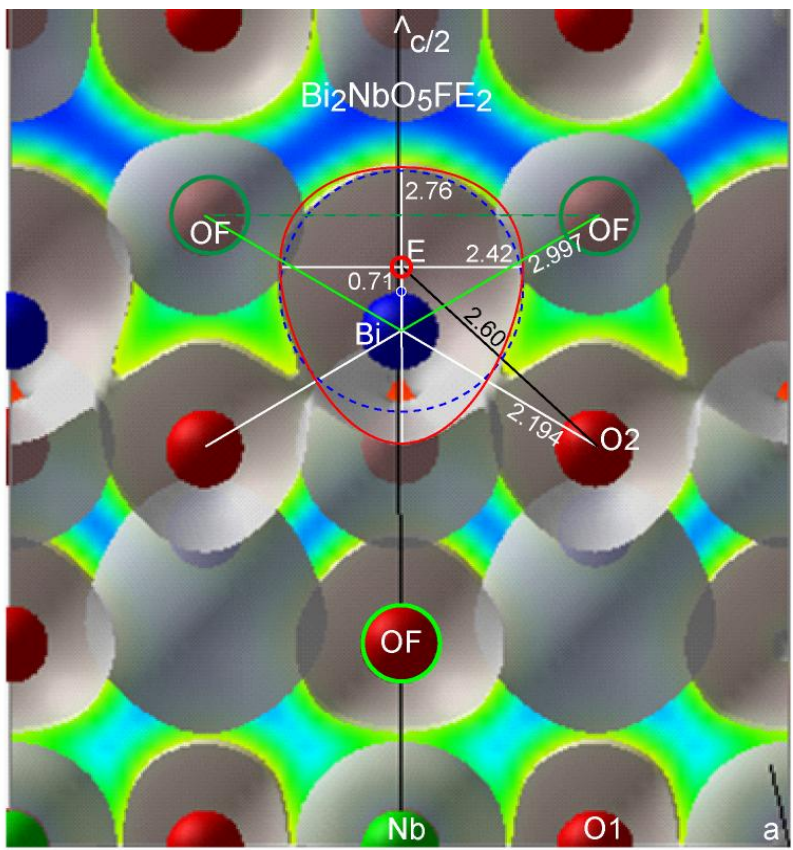

a)

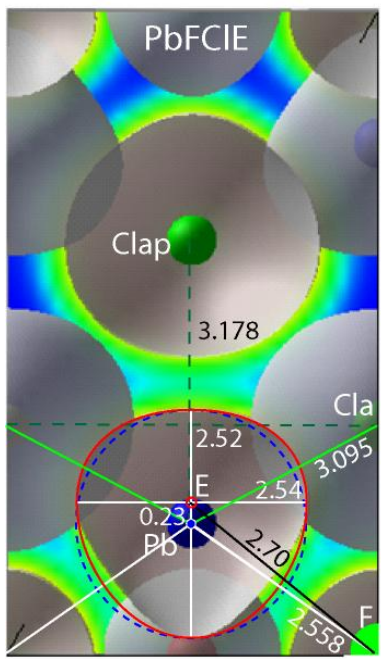

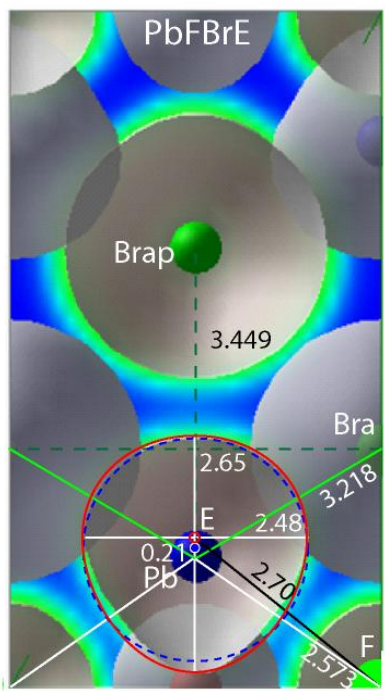

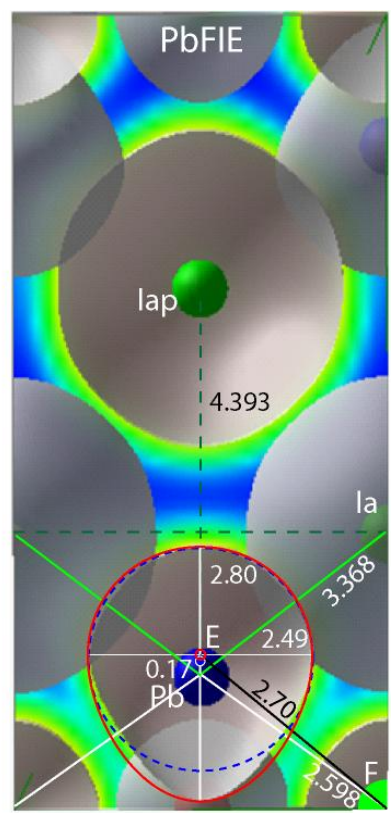

b)

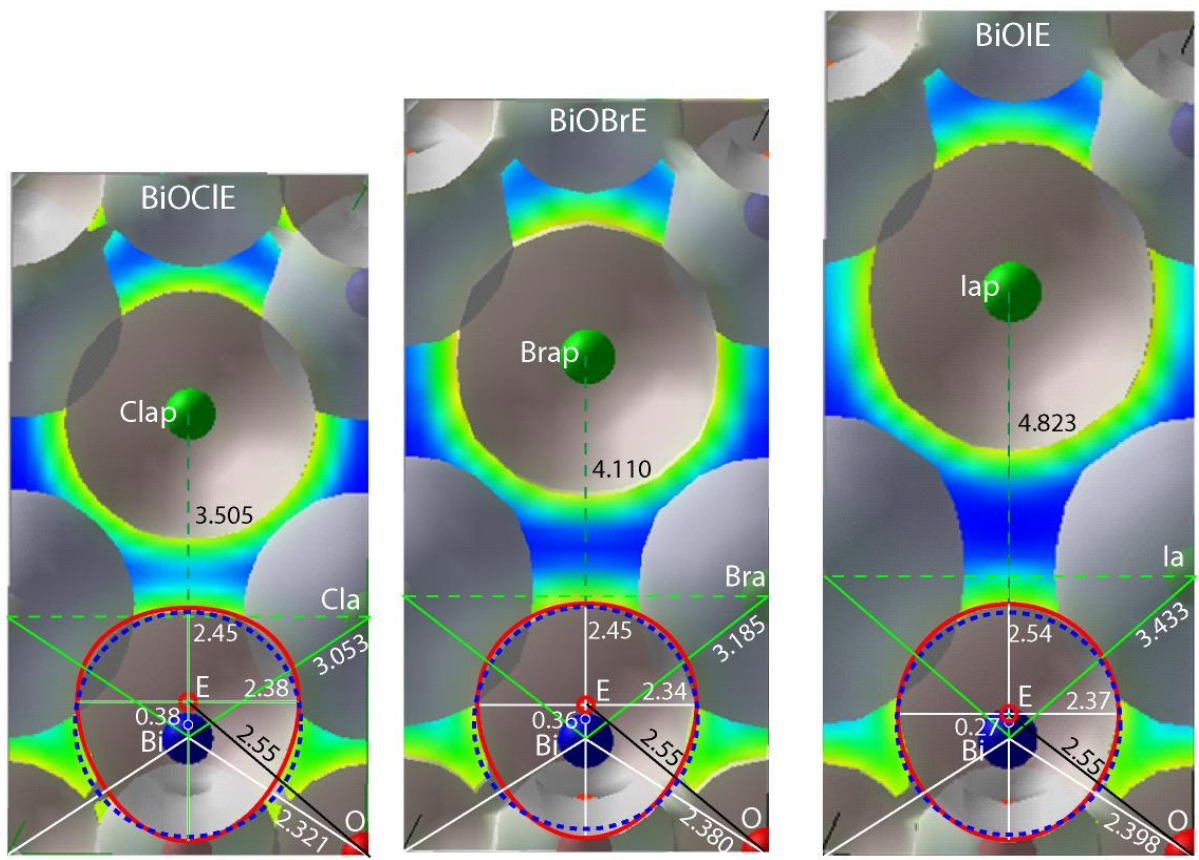

c)

**Figure 8** (color) ELF 3D isosurface grey shells with configurational description of the LP (see text) in a) BiOF with $BiE_2NbO_5F$, b) PbFX series: c) BiOX series. The background ELF projections (2D slices) are shown with dominant zero localization between the atomic constituents.

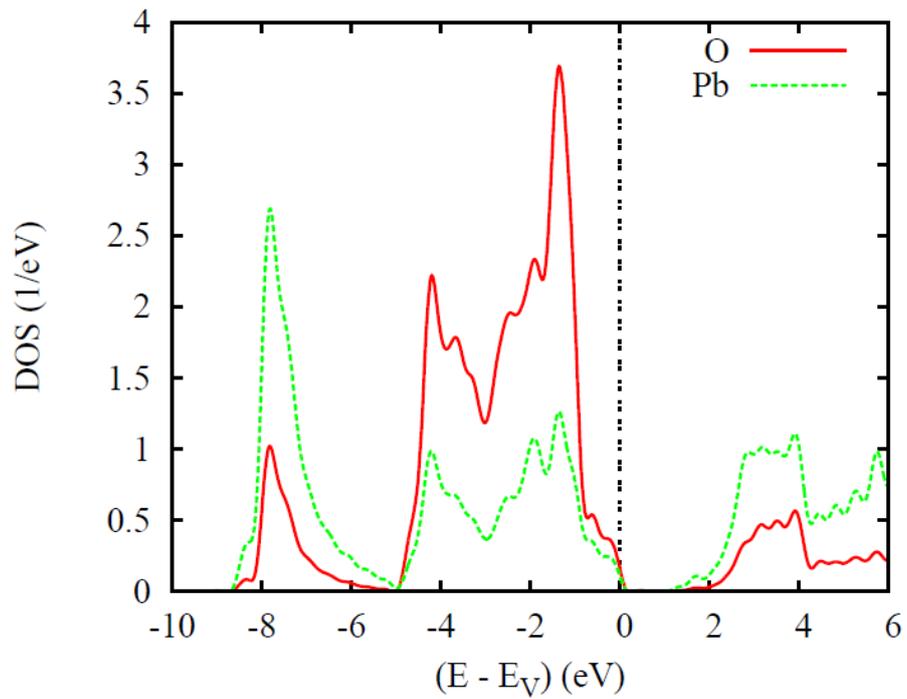

a)

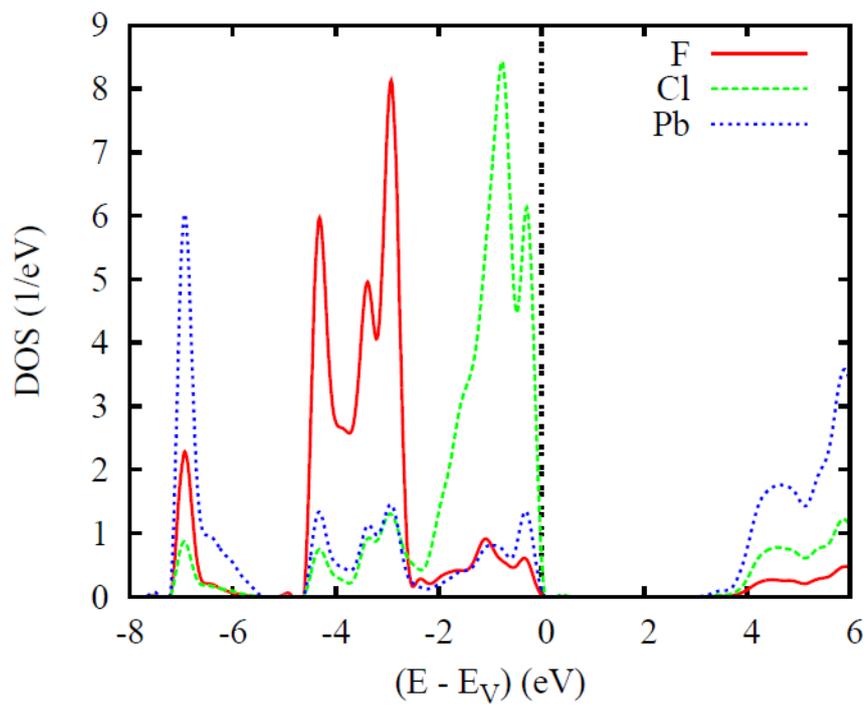

b)

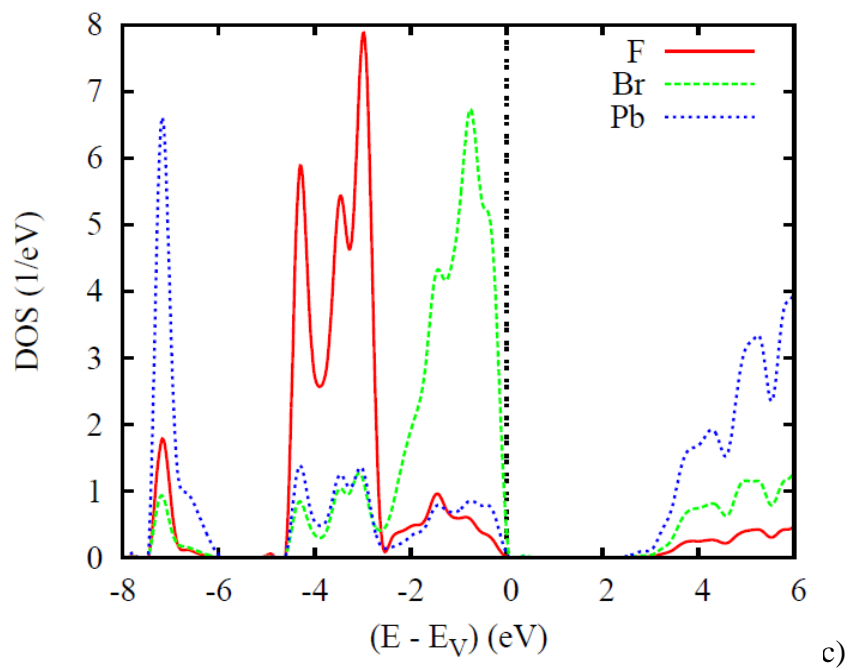

c)

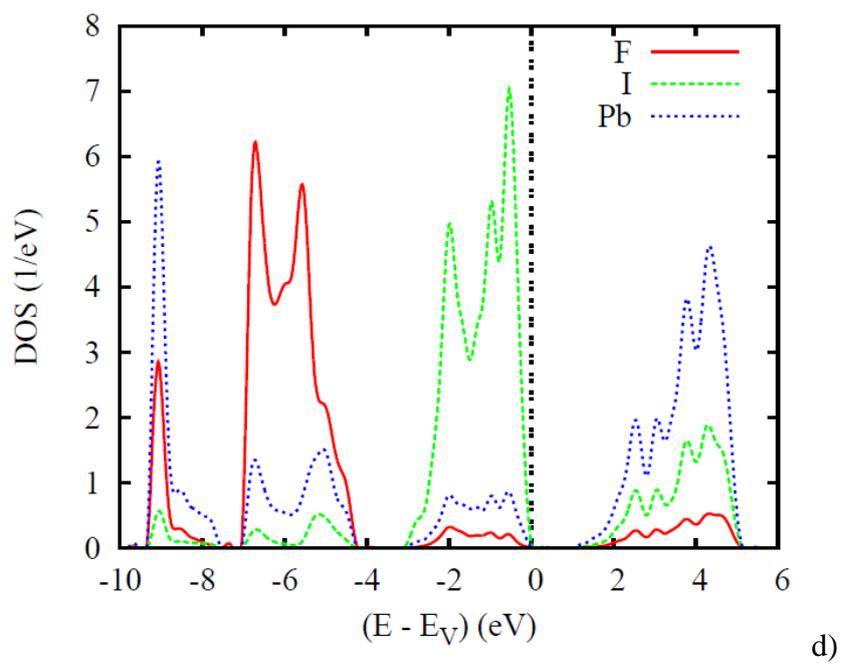

d)

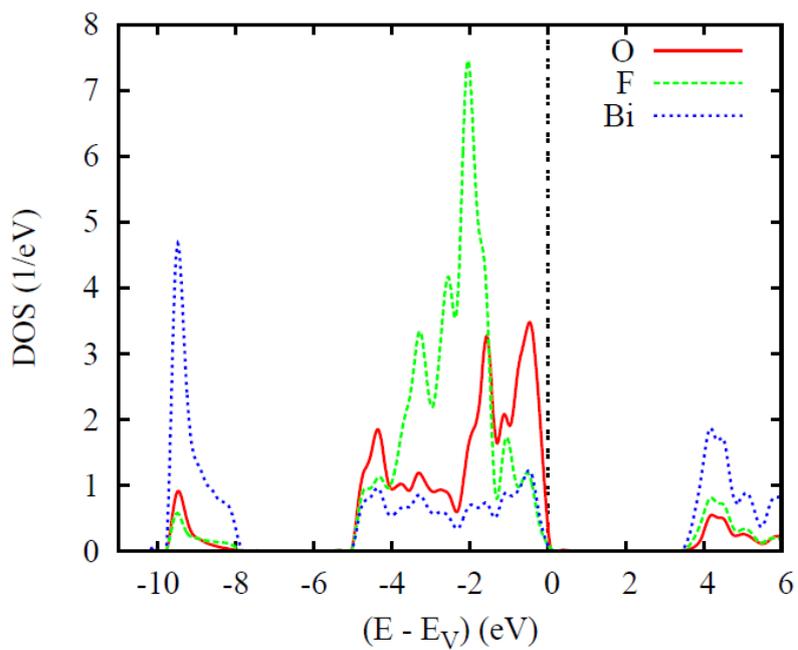

e)

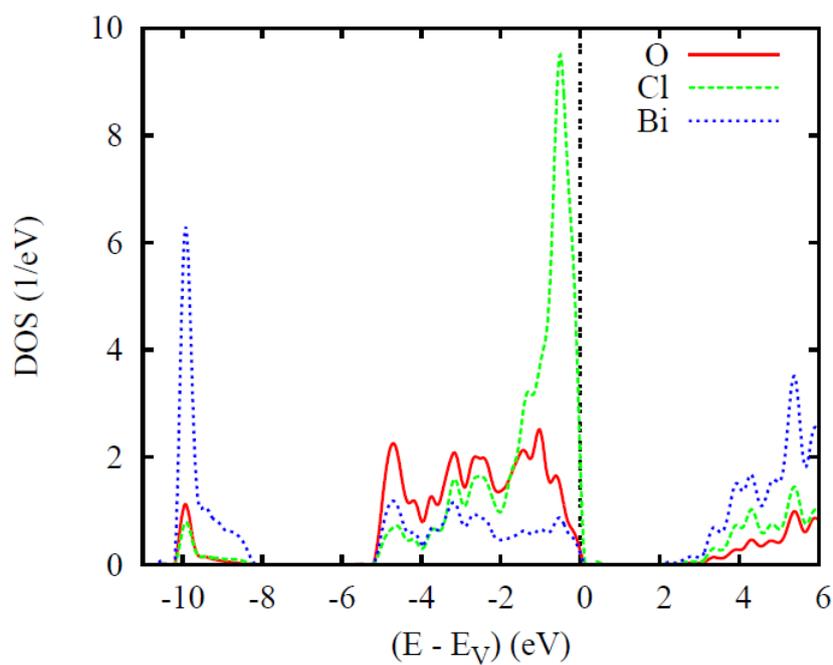

f)

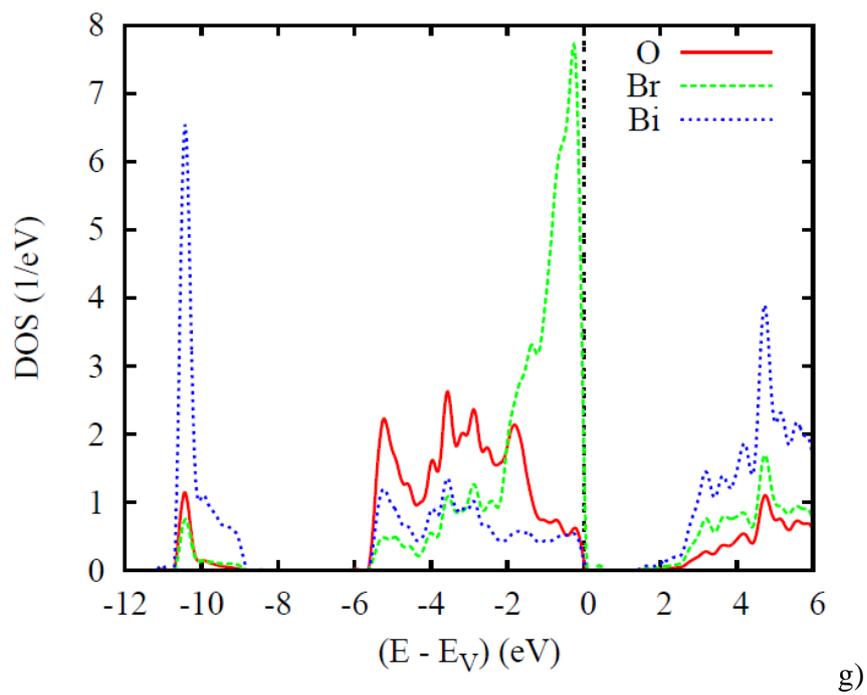

g)

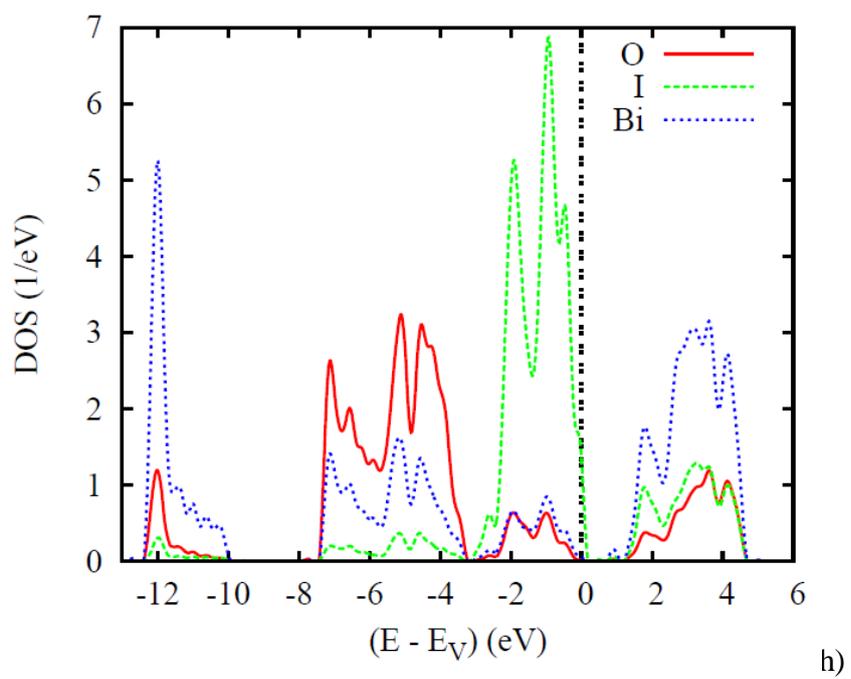

h)

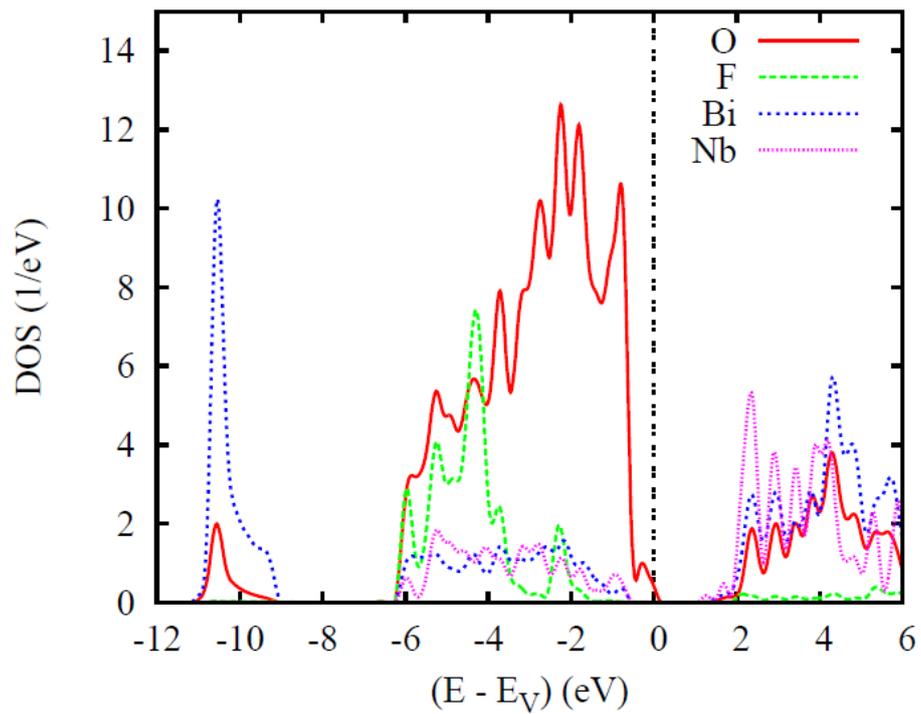

i)

**Figure 9** (color) Site projected density of states (DOS) a) PbO, b-d) PbFX family, e – h) BiOX family and i) $Bi_2NbO_5F$.

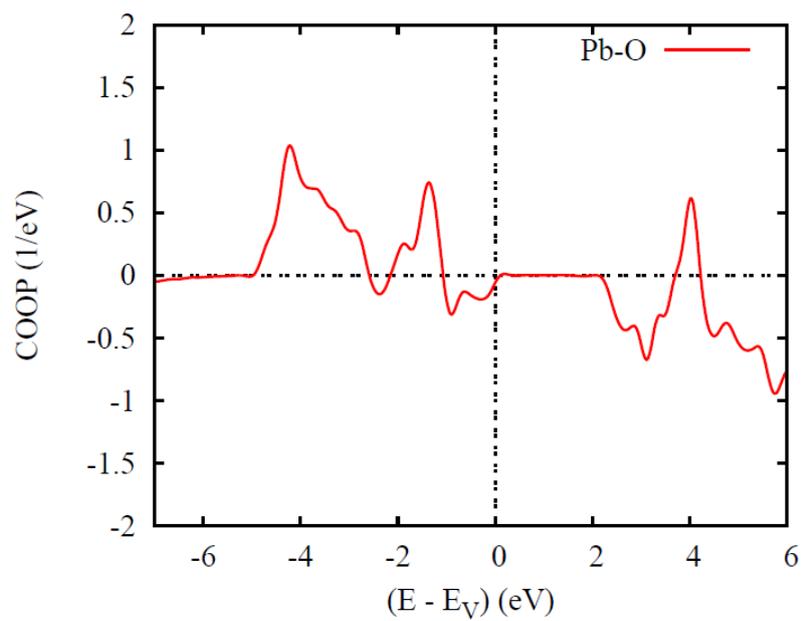

a)

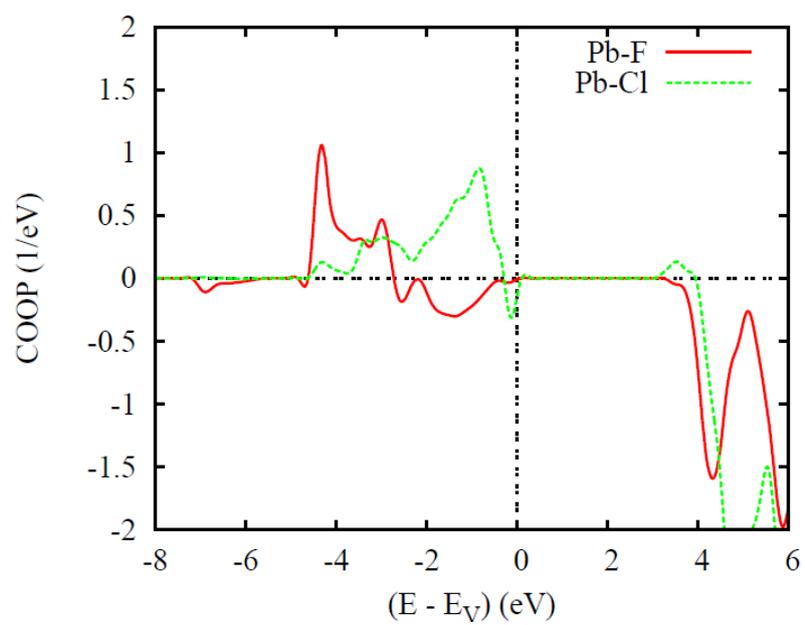

b)

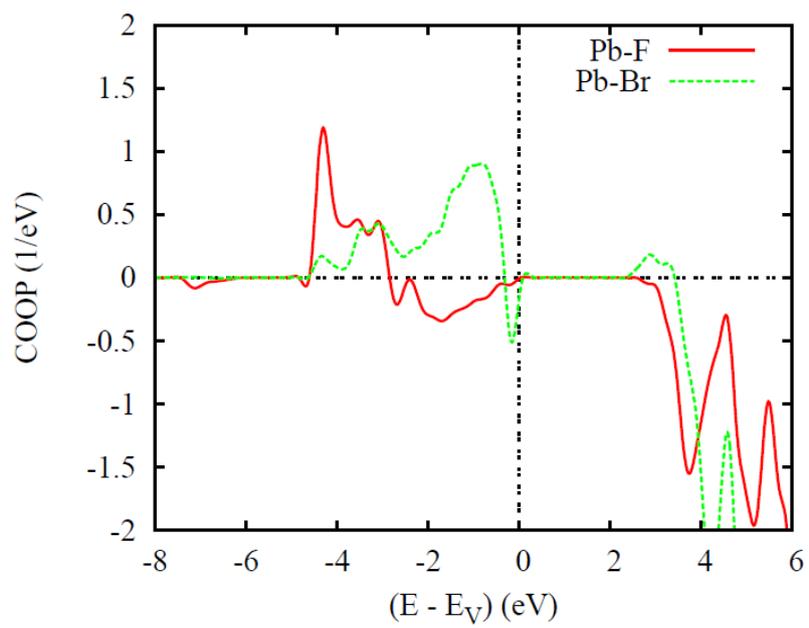

c)

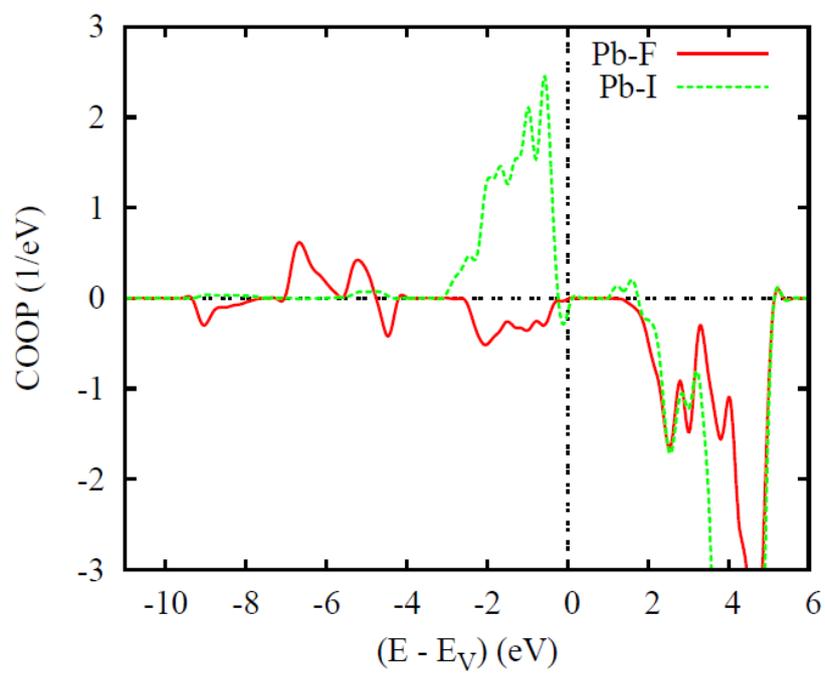

d)

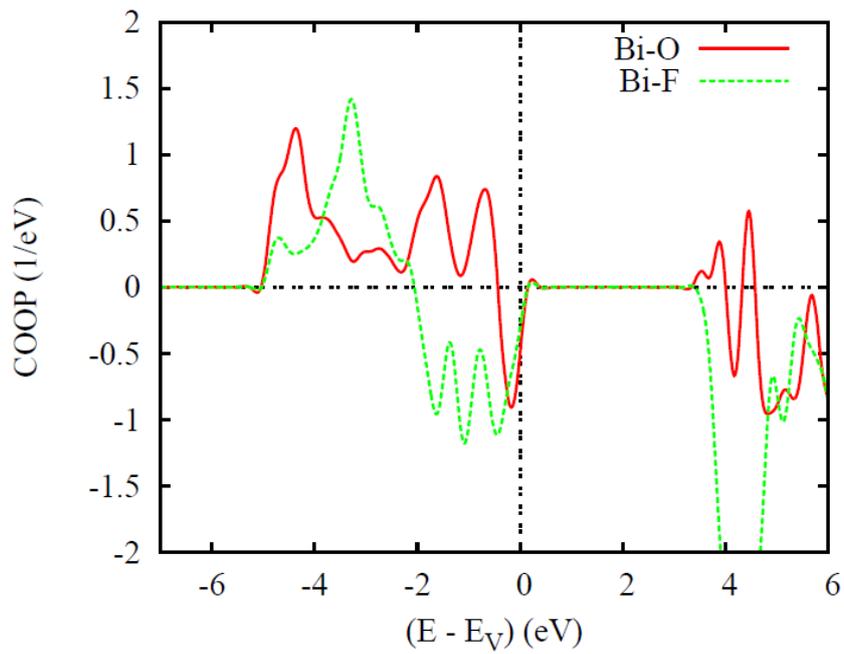

e)

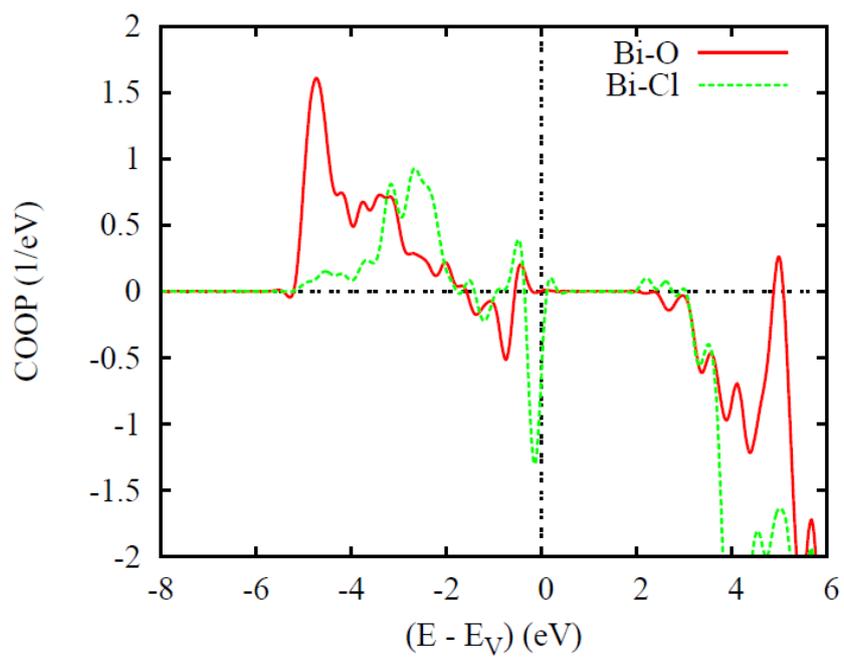

f)

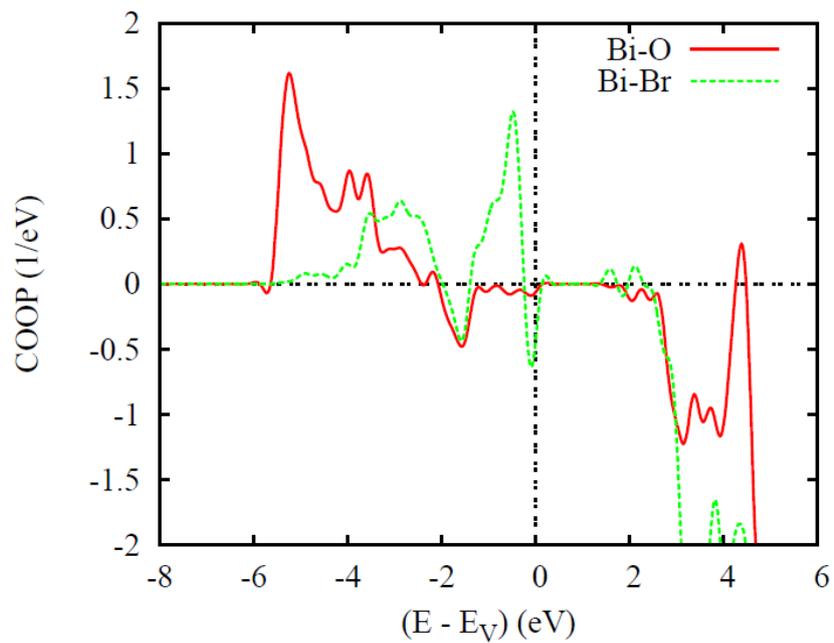

g)

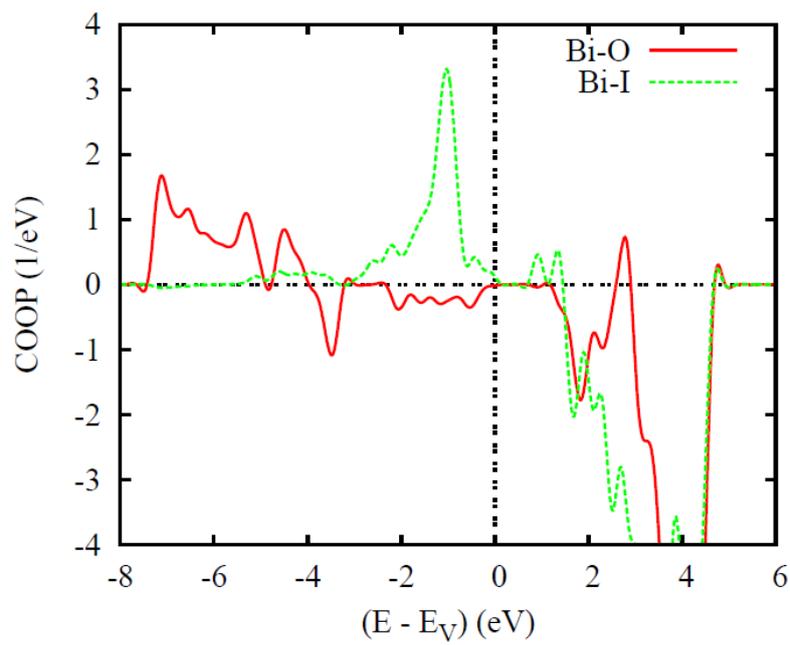

h)

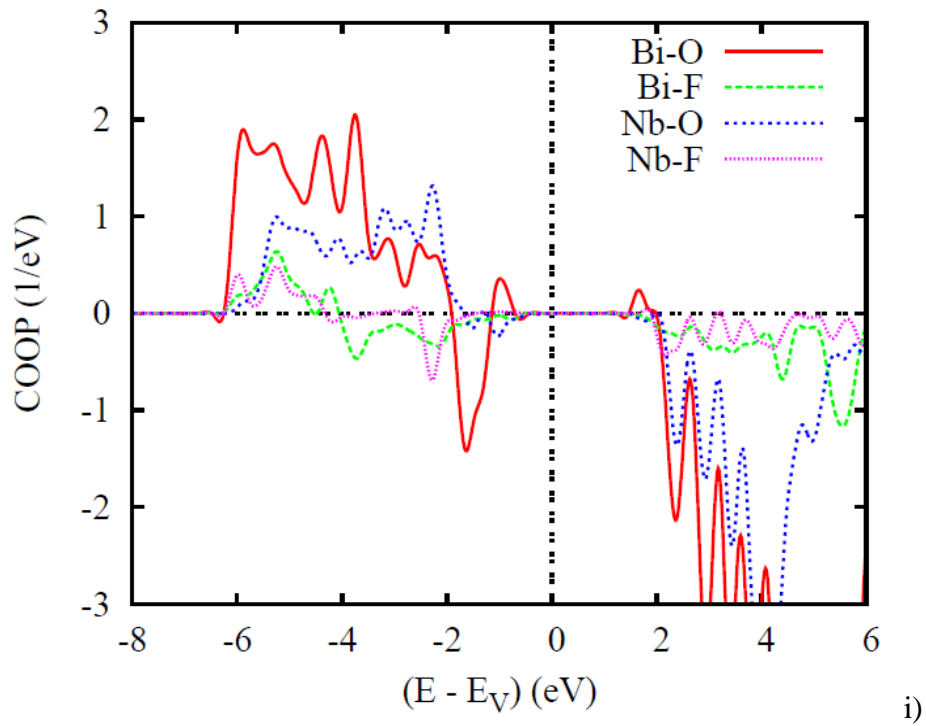

**Figure 10** (color) Chemical bonding for pair inteactions with COOP criterion in: a) PbO, b) – d) PbFX family, e) – h) BiOX family and i) $Bi_2NbO_5F$.